\documentclass[lettersize,journal]{IEEEtran}
\usepackage{amsmath,amsfonts}
\usepackage{algorithmic}
\usepackage{algorithm}
\usepackage{color}
\usepackage{array}
\usepackage[caption=false,font=footnotesize,labelfont=rm,textfont=rm,subrefformat=parens]{subfig}
\DeclareSubrefFormat{parens}{#1(#2)}
\usepackage{textcomp}
\usepackage{stfloats}
\usepackage{url}
\usepackage{bm}
\usepackage{verbatim}
\usepackage{graphicx}
\usepackage{cite}
\usepackage{amsmath}
\usepackage{autobreak}
\usepackage{makecell,rotating,multirow,diagbox}
\usepackage[mathscr]{euscript}
\usepackage{graphicx}
\usepackage{float}
\usepackage{subfig}
\usepackage{booktabs}

\makeatletter
\newcommand{\removelatexerror}{\let\@latex@error\@gobble}
\makeatother
\def\BibTeX{{\rm B\kern-.05em{\sc i\kern-.025em b}\kern-.08em
		T\kern-.1667em\lower.7ex\hbox{E}\kern-.125emX}}
\usepackage{balance}
\begin{document}
	\title{Integrated Sensing and Communications Towards Proactive Beamforming in mmWave V2I via Multi-Modal Feature Fusion (MMFF)}
	\author{Haotian Zhang,~\IEEEmembership{Graduate Student Member,~IEEE,}  Shijian Gao,~\IEEEmembership{Member,~IEEE,} Xiang Cheng,~\IEEEmembership{Fellow,~IEEE,} Liuqing Yang,~\IEEEmembership{Fellow,~IEEE,}
		
		% <-this % stops a space
		\thanks{Haotian Zhang and Xiang Cheng are with the State Key Laboratory of Advanced Optical Communication Systems and Networks, School of Electronics, Peking University, Beijing 100871, China (e-mail: haotianzhang@stu.pku.edu.cn; xiangcheng@pku.edu.cn).
			
		Shijian Gao is with the Internet of Things Thrust, The Hong Kong University of Science and Technology (Guangzhou), Guangzhou 511400, China (e-mail: shijiangao@hkust-gz.edu.cn).
		
		Liuqing Yang is with the Internet of Things Thrust and Intelligent Transportation Thrust, The Hong Kong University of Science and Technology (Guangzhou), Guangzhou 511400, China, and also with the Department of Electronic and Computer Engineering and the Department of Civil and Environmental Engineering, The Hong Kong University of Science and Technology, Hong Kong, SAR, China (e-mail: lqyang@ust.hk).
		}% <-this % stops a space
	}
	
%	\markboth{IEEE TRANSACTIONS ON WIRELESS COMMUNICATIONS,~Vol.~XXX, No.~XXX, XXX~2023}%
%	{How to Use the IEEEtran \LaTeX \ Templates}
	
	\maketitle
	
	\begin{abstract}
	
	The future of vehicular communication networks relies on mmWave massive multi-input-multi-output antenna arrays for intensive data transfer and massive vehicle access. However, reliable vehicle-to-infrastructure links require {\color{black} exact alignment between the narrow beams}, which traditionally involves excessive signaling overhead. To address this issue, we propose a novel proactive beamforming scheme that integrates multi-modal sensing and communications via Multi-Modal Feature Fusion Network (MMFF-Net), which is composed of multiple neural network components with distinct functions. Unlike existing methods that rely solely on communication processing, our approach obtains comprehensive environmental features to improve beam alignment accuracy. We verify our scheme on the {\color{black}Vision-Wireless (ViWi)} dataset, which we enriched with realistic vehicle drifting behavior. Our proposed MMFF-Net achieves more accurate and stable angle prediction, which in turn increases the achievable rates and reduces the communication system outage probability. Even in complex dynamic scenarios {\color{black}with adverse environment conditions}, robust prediction results can be guaranteed, demonstrating the feasibility and practicality of the proposed proactive beamforming approach.

	\end{abstract}
	

	\section{Introduction}
	%第一段：背景,车联网日益重要以及车联网的通信系统采用毫米波频段,相应带来波束对准的问题
	\IEEEPARstart{T}{he} vehicular communication network (VCN) is playing an increasingly important role in intelligent transportation systems \cite{xu2017internet,lu2014connected,cheng2022integrated}. To meet the ultra-reliable and low latency communication requirements of VCN applications \cite{chen2017vehicle, cheng2022mmwave}, millimeter-wave (mmWave) communications have been recognized as one of the key enablers for {\color{black}the fifth-generation (5G)} and beyond wireless systems \cite{boccardi2014five,rusek2012scaling,mishra2020toward}. To compensate for the high path loss in the mmWave band and reduce interference to non-target users \cite{el2014spatially, gao2021mutual,eldar2022interference}, the transmitter typically deploys a massive multi-input-multi-output (mMIMO) antenna array and formulates ``pencil-like" beams to focus energy on the directions of users. Due to the narrow beamwidth, it is crucial to align the transmit and receive beams between the road side unit (RSU) and vehicles to establish a reliable vehicle-to-infrastructure (V2I) communication. 
	
	%第二段：传统的波束对准方案都是基于comm-only
	Conventional beam alignment methods generally design the precoding matrix based on dedicated beam training \cite{wang2009beam, xiao2016hierarchical,  va2016beam,alkhateeb2014channel,gao2020estimating}. The typical solution involves the exhaustive search for the beam pair with the strongest signal-to-noise ratio (SNR) in the codebook \cite{wang2009beam, xiao2016hierarchical}. However, the subsequent high overhead and latency make it impractical for VCN, where the beamforming angle needs to be updated frequently due to potentially frequent channel variations \cite{myCOMST}. To overcome this deficiency, \textit{beam tracking} \cite{va2016beam} has been proposed to reduce the search space of the optimal beam pairs by utilizing temporal correlation. However, the periodic signaling interaction is still inevitable. To further reduce the pilots, \textit{proactive beamforming} has recently been proposed to avoid frequent training from the scratch. Existing proactive beamforming schemes generally adopt simple kinematic models and take up additional spectrum resources to conduct motion detection or state feedback \cite{shaham2019fast,liu2020radar,yuan2020bayesian,yuan2021integrated,mu2021integrated, liu2022learning}. They mainly rely on state evolution models to track the vehicle's motion parameters, such as the {\color{black}Kalman filter (KF) or extended Kalman filter (EKF)} \cite{shaham2019fast,liu2020radar}, which limits their applicability to ideal straight paths. To make the application scenarios more generic, Liu \textit{et al.}~\cite{liu2020tutorial} proposed a maximum likelihood estimator-based scheme using the vehicle's historical trajectory data rather than a state evolution model. However, the above methods are subject to approximations due to the high nonlinearity of the state evolution model, which limits the prediction accuracy.
	
	To address this limitation, researchers have started exploring alternative methods that rely on deep learning (DL) techniques. Specifically, in \cite{mu2021integrated, liu2022learning}, DL has been utilized to vehicle's motion prediction and beamforming matrix prediction with received signals and historical estimated channels serving as input. Though the above schemes partially address the problem of predicting the vehicle's angular parameter, they primarily adopt the electromagnetic propagation environment information to learn the vehicle's motion parameter evolution characteristic. This limitation restricts the application of such schemes in complex and dynamic environments.
	
	Fortunately, with the increasing prevalence of advanced sensors in intelligent transportation infrastructure, various studies have explored the benefits of incorporating multi-modal sensing into communication system design \cite{ alrabeiah2020millimeter,demirhan2022radar,klautau2019lidar,wu2022lidar,charan2021vision, fan2022radar,differentiation_of_multi_v}. This aligns with a crucial objective of the integrated sensing and communications (ISAC), aiming to achieve mutual enhancement between communications and sensing. This includes communication-assisted sensing and sensing-assisted communications, operating at a level beyond hardware or waveform considerations \cite{LiuJSAC}. Within the category of sensing aided communications, Alrabeiah \textit{et al.} \cite{alrabeiah2020millimeter} proposed to employ red-green-blue (RGB) images to aid mmWave beam and blockage prediction tasks in VCN. Demirhan \textit{et al.} \cite{demirhan2022radar} developed beam prediction schemes based on radar raw data, range-angle maps, and range-velocity maps. In \cite{klautau2019lidar} and \cite{wu2022lidar}, the light detection and ranging (LiDAR) point cloud was leveraged for beam selection and link blockage prediction tasks. Charan \textit{et al.} \cite{charan2021vision} combined sequences of images and mmWave beams to predict blockages and proactively initiate hand-off decisions.

	The work discussed in   \cite{alrabeiah2020millimeter,demirhan2022radar,klautau2019lidar,wu2022lidar,fan2022radar,charan2021vision, differentiation_of_multi_v} uses uni-modal sensory data, which falls short in offering comprehensive environmental features and yields limited gains in communication functionalities.  Furthermore, the non-radio frequency  sensors are sensitive to weather and lighting conditions, causing uni-modal sensor-based schemes to fail in certain cases.  Recently, Cheng \textit{et al.} \cite{SoM} raised that multi-modal sensing can potentially reflect the channel propagation characteristics and bring broader functional optimization in communication systems beyond hardware resources, which is summarized as the Synesthesia of Machines (SoM) framework. However, merely combining multi-modal data is far from sufficient in real use, thus advanced data processing and feature analysis require an in-depth investigation. Firstly, efficient data pre-processing methods for multi-modal data with different semantic information and data structures must be developed to extract relevant features and eliminate redundant information. Secondly, dedicated DL models capable of extracting multi-modal features and reasonably fusing them need to be explored. To this end, we propose a Multi-Modal Feature Fusion Network (MMFF-Net) which includes three judiciously designed feature extraction modules, {\color{black}an intelligent multi-modal feature fusion mechanism}, and two distinct regression models for predicting the beamforming angle, {\color{black}realizing the goal of sensing-assisted communications}. Our main contributions are summarized as follows:
	
	\begin{itemize}
		\item[ $\bullet$]
		{\color{black}The proposed scheme addresses the V2I proactive beamforming issue through a \textbf{dedicated neural network (NN) construction}, essentially realizing the integration of multi-modal sensing and communications. This approach incorporates multi-modal feature extraction and fusion modules with adaptive weight learning to intelligently obtain the motion characteristics of a vehicle.} The MMFF-Net includes two regression models to accommodate the high-dimensional input and the vehicle's random drifting behavior, aiming at accurately predicting the vehicle's future position. The proposed scheme is suitable for scenarios where the vehicle drifts randomly while driving.
	\end{itemize}

	\begin{itemize}
		\item[ $\bullet$]
		The dataset is enriched with complex user behavior to expand the applicability of proactive beamforming. It is constructed based on the {\color{black}Vision-Wireless (ViWi)} data-generation framework \cite{alrabeiah2020viwi} and further enhanced by considering the vehicle's \textbf{random drifting behavior}. Each trajectory point in the dataset is accompanied by synchronized RGB images, depth maps, and {\color{black}channel state information (CSI)}. The proposed MMFF-Net has been proven to efficiently support more complicated and realistic scenarios. {\color{black}Furthermore, the robustness of MMFF-Net against environmental interference in practical applications is demonstrated through validating the trained network model on datasets added with noises.}
	\end{itemize}

	\begin{itemize}
	\item[ $\bullet$]    
		The proposed scheme outperforms up to \textbf{eight} benchmark schemes in terms of the prediction accuracy and adaptability to the vehicle's behavior. MMFF-Net reduces the angle prediction error by \textbf{31.52 percent} in the overall process and by \textbf{11.76 percent} when vehicle approaches the RSU in comparison with the benchmark schemes.
%		, and {\color{red}17.62 percent} when the vehicle approaches RSU and relative angle varies fast. ekf-t:0.0194 image:0.0199 our:0.015
		This in turn leads to \textbf{an achievable rate gain of at least 25.30 percent} and \textbf{an outage probability reduction of at least 15.69 percent} when a large antenna array is adopted. %Additionally, the proposed scheme minimizes the outage probability of the communication system when adopted. 
		The proposed algorithm will be made open source, providing a reference point for future research.
		
		%32 antenna i-pb 6.64 our:8.32
	\end{itemize}

	The remainder of this paper is organized as follows. Section \ref{problem} introduces the scenario considered and the proactive beamforming problem to be solved. Section \ref{Extract} presents the operation process of the multi-modal feature extraction modules. Section \ref{IV} details the multi-modal feature fusion and beamforming angle prediction. Section \ref{V} provides the details of the experimental setup to verify the proposed scheme. Section \ref{Performance} presents extensive simulation results and analyses. The conclusions are given in Section \ref{VI}.
	
	\textit{Notations:} Throughout this paper, we use a capital boldface letter $\textbf{A}$, a lowercase boldface letter $\textbf{a}$, and a lowercase letter $a$ to represent a matrix, a vector, and a scalar, respectively. $\textbf{A}[m,:]$ and $\textbf{A}[:,m]$ represent the $m$-th row and $m$-th column of $\textbf{A}$. $\mathbb{C}^{M\times N}$ represents a complex space of dimension $M \times N$ and $\mathbb{R}^{M\times N \times K}$ represents a real space of dimension $M \times N \times K$. $\textbf{A}^{-1}$, $\textbf{A}^{\rm{T}}$, $\textbf{A}^{\rm{H}}$, and $\textbf{A}^{\dagger}$ represent the inverse, transpose, Hermitian, and pseudo-inverse transpose of $\textbf{A}$, respectively. $[a]$ represents the operation of rounding $a$ to the nearest whole number. $\lvert \cdot \rvert$ represents the modulus of a complex number. $\Vert \cdot \Vert_{\rm{F}}$ represents the Frobenius norm of a matrix. The notations $\otimes$ and $\oplus$ represent the matrix element-wise product and the concatenation operation, respectively. Given vectors $\textbf{a} \in \mathbb{C} ^{1\times M}$ and $\textbf{b} \in \mathbb{C} ^{1\times N}$, $\textbf{a} \oplus \textbf{b} = [\textbf{a},\textbf{b}] \in \mathbb{C} ^{1\times (M+N)}$. 
	
	\section{Scenario and Problem Description}
	\label{problem}
	In this section, we provide a high-level overview of the VCN scenario studied in this paper. Subsequently, we describe the proactive beamforming problem that we aim to address.
	\subsection{Scenario}
	\label{scenario}
\begin{figure}[!t]
		\centering
		\includegraphics[width=1\linewidth]{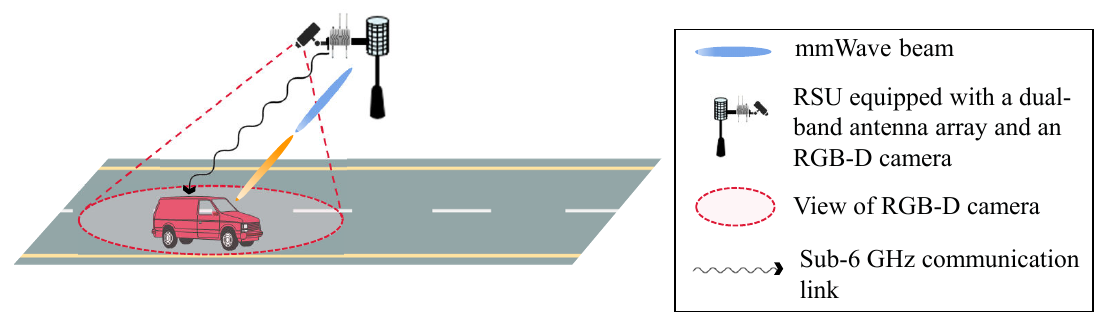}
		\caption{An illustration of the considered V2I scenario model.}
		\label{f1}
	\end{figure}

	In this paper, we consider a VCN scenario where an RSU is serving a passing vehicle. We assume that the C/U-plane decoupled network architecture is adopted, which integrates sub-6 GHz bands and broadband mmWave bands to expand the system bandwidth. The control channel is assumed to operate at sub-6 GHz bands and transmit control information between the vehicle and RSU. The data channel is assumed to operate at mmWave bands and transmit user data. As depicted in Fig. \ref{f1}, the RSU is equipped with a MIMO uniform linear array (ULA) which operates at both mmWave band and sub-6 GHz band with $N^{\rm{HF}}_{\rm{t}}$ transmit antennas at mmWave band, and $N^{\rm{LF}}_{\rm{t}}$ transmit antennas at sub-6 GHz band. The vehicle is assumed to be equipped with a MIMO ULA which also operates at both the mmWave band and sub-6 GHz band with $M^{\rm{HF}}_{\rm{r}}$ receive antennas at mmWave band and single antenna at sub-6 GHz band. The communication system is assumed to adopt the orthogonal frequency division multiplexing (OFDM) technique with $K_{\rm{HF}}$ subcarriers at the mmWave band and $K_{\rm{LF}}$ subcarriers at the sub-6 GHz band. In order to collect real-time multi-modal sensory data, the RSU is also equipped with a {\color{black}red-green-blue-depth (RGB-D)} camera. Note that the depth data (the distances from the object to the depth camera) added with Gaussian noise is used to mimic frequent radar sensing.

	\subsection{Problem Description}
	%In order to achieve ultra-reliable communications in latency-critical VCNs, it is crucial for RSUs to acquire accurate relative angles of passing vehicles in real time, and formulate a narrow transmit beam that accurately points to the vehicle. 
	\label{formulation}
	To avoid large signaling overhead and latency caused by beam training or tracking methods, RSUs can proactively predict the beamforming angle in the next time slot to formulate a narrow transmit beam that accurately points to the vehicle.
	
    We propose to adopt multi-modal sensory data and wireless channel data to assist the RSU in achieving the above goal. Specifically, RSU senses the traffic environment at a fixed frequency and communicates with the passing vehicle constantly. At the $(n-1)$-th time slot, RSU predicted the relative azimuth angle $\hat{\theta}_n$ of the passing vehicle, which is also the angle of departure (AoD) of the transmitted signal, using the sequences of RGB images, depth maps, and CSI of sub-6 GHz channel. The actual AoD is denoted by $\theta_n$. To achieve this, dedicated pre-processing methods for multi-modal data and feature fusion based on DL model need to be developed. Note that the vehicle also needs to formulate the narrow beam that points to RSU based on the angle of arrival (AoA) of the signal at the vehicle's antenna array when it adopts a MIMO ULA, which is denoted by $\phi_n$. For simplicity but without loss of generality, we assume the vehicle drives parallel to the RSU's antenna array, {\color{black} implying $\theta_n=\phi_n, {\forall} n$.}\footnote{{\color{black}This assumption may not hold in some practical road scenarios. However, it's important to note that our study focuses on how the RSU can use multi-modal environmental data to assist proactive beamforming. This assumption is made for the convenience of validating the accuracy of beamforming angles predicted by the RSU. When this assumption does not hold, the AoA at the vehicle can still be easily estimated using array signal processing algorithms such as Multiple Signal Classification (MUSIC), as shown in \cite{liu2020tutorial,MUSIC1}, without affecting the implementation of MMFF-Net at the RSU.}} Therefore, we omit $\phi_n$ in this paper and assume that RSU conducts the angle prediction and sends the result to the vehicle instantly within the current downlink transmission block to ensure the link establishment at the next time instance. With the angle prediction result $\hat{\theta}_n$, both the RSU and vehicle accordingly formulate the narrow beams using the transmit beamforming matrix $\mathbf{f}_n= \mathbf{a}(\hat{\theta}_n)$ and receive beamforming matrix $\mathbf{w}_n=\mathbf{b}(\hat{\theta}_n)$, respectively. $\mathbf{a}(\theta)$ and $\mathbf{b}(\theta)$ are the transmit steering vector of the RSU's antenna array and receive steering vector of the vehicle's antenna array, respectively. The goal of a desirable proactive beamforming scheme is to predict the future relative angle as accurately as possible and maintain stable results when facing realistic random behavior, thus maximizing the achievable rate and reducing the outage probability of the communication system.  This could be mathematically expressed as\footnote{Detailed description of function ${\rm SNR}_n(\cdot)$ will be given in Section \ref{Signal model}.}
    \begin{equation}
    \begin{aligned} \label{Object}
        &\mathop{\max}\limits_{\mathbf{\Theta}} \ R_n(\mathbf{\Theta}) =  \log_2(1+{\rm SNR}_n(\hat{\theta}_n)) \\
        & s.t. \quad \hat{\theta}_n = f_{\mathbf{\Theta}}(\mathbf{h}_{n-2}^{\rm{P}'},\mathbf{h}_{n-1}^{\rm{P}'},\hat{\mathbf{X}}^{\rm{P}}_{n-2},\hat{\mathbf{X}}^{\rm{P}}_{n-1},\hat{\mathbf{Y}}^{\rm{P}}_{n-2},\hat{\mathbf{Y}}^{\rm{P}}_{n-1})
    \end{aligned} 
    \end{equation}
    where $f_{\mathbf{\Theta}}(\cdot)$ is the proposed MMFF-Net model; $\mathbf{\Theta}$ is the set of model parameters; $\mathbf{h}_{n-1}^{{\rm{P}}'}$, $\hat{\mathbf{X}}^{\rm{P}}_{n-1}$, and  $\hat{\mathbf{Y}}^{\rm{P}}_{n-1}$ are the ultimate pre-processed CSI matrix, RGB image, and depth map at the $(n-1)$-th time instance, respectively.

	\section{Multi-Modal Feature Extraction}
	\label{Extract}
	While serving the passing vehicle,  RSU continuously senses the traffic environment. At each  slot, it proactively predicts the future position of the vehicle and optimizes beamforming using multi-modal sensory data and wireless channel data. However, multi-modal data has distinct semantic information and data formats, which calls for dedicated data pre-processing and feature extraction that relate to vehicle motion parameters. To address this, we design a novel MMFF-Net consisting of three distinct feature extraction modules, each utilizing different data pre-processing methods and containing NNs with distinct architectures and functions. For simplicity, in the following content, all subscripts $n$ refer to certain data at the $n$-th time instance and will not be reiterated.
	
	\subsection{Angular Feature Extraction Module}
    \label{AFE} 

    CSI can be regarded as the compressed electromagnetic environment characterization, which is roughly characterized by the relative angle at the passing vehicle. Therefore, we define the angular feature related to the line-of-sight (LoS) channel's azimuth. Benefiting from the angular feature, the NN can learn the vehicle's motion state evolution characteristics more efficiently and predict the future position more accurately.
	
	However, obtaining the angular feature faces challenges. First, the MIMO and OFDM techniques jointly lead to a large CSI matrix which causes the high processing load of NN and long training time. Second, the acquisition of CSI in mmWave bands requires excessive signaling overhead and time to transmit sufficient training symbols due to the limited radio frequency (RF) chains in the commonly adopted hybrid structure. Third, raw CSI does not explicitly reflect the angular information of the signal transmission paths though it is a joint result of them. In response to these challenges, we design an angular feature extraction (AFE) module to pre-process the raw CSI into a low-dimensional one and extract the implied angular feature. To avoid the excessive signaling overhead and latency, we assume that RSU obtains the CSI through necessary control signaling (such as synchronous signal) transmitted at sub-6 GHz bands. Though MIMO beamforming is conducted at mmWave bands, CSI at lower frequencies still reflects extremely similar signal propagation process with that of mmWave bands, especially the LoS component \cite{submmwave-4,mmMAGIC}. Existing studies \cite{submmwave-1,submmwave-2,gao2021fusionnet} also prove such a correlation by predicting the optimal mmWave downlink beam directly from sub-6 GHz channels. Moreover, the co-existing sub-6 GHz and mmWave antenna structure has also appeared in academia and industry \cite{submmwave-3}. Considering these facts, we propose to use CSI of sub-6 GHz bands to provide rough angular information of vehicle in the electromagnetic domain. In what follows, we will introduce the pre-processing methods of raw CSI and the extraction of the angular feature.
	
	\subsubsection{Pre-Processing of Raw CSI}
	\label{preprocess}
	The proposed CSI pre-processing method is dedicated to transforming the primitive CSI matrix into a representation with explicit angular features, which is composed of three steps.

	\textbf{\textit{Step 1: Beam energy calculation and dimensionality reduction}}. The size of sub-6 GHz antenna arrays is typically small. In this case, using a conventional discrete Fourier transform (DFT) codebook to calculate the energy of CSI at different beam angles can only provide extremely rough directional information due to the limited number of steering vectors therein. Therefore, we adopt a super-resolution DFT codebook $\mathbf{D}=\left\{\mathbf{d}_1,\mathbf{d}_2,...,\mathbf{d}_{\rm{B}}\right\}$ to calculate the total energy of all subcarriers at each beam angle, where $\mathbf{d}_{\rm{b}} \in \mathbb{C}^{N^{\rm{LF}}_{\rm{t}}\times 1}$ and $B$ denote the $b$-th steering vector and angle resolution, respectively. Let $\mathbf{H}_n[k] \in \mathbb{C}^{N^{\rm{LF}}_{\rm{t}}\times 1}$ be the sub-6 GHz channel at the $k$-th subcarrier, and $\mathbf{h}_n^{\rm{P}} \in \mathbb{R}^{B\times 1}$ be the processed CSI matrix after \textit{energy calculation} and \textit{dimensionality reduction}. Then, $\mathbf{h}_n^{\rm{P}}[i]$ is represented as:
	\begin{equation}
		\label{energy}
	  \mathbf{h}_n^{\rm{P}}[i] = \frac{1}{K_{\rm{LF}}}\sum_{k=1}^{K_{\rm{LF}}}\log_{2}{(1+\rho \lvert \mathbf{d}_i^{\rm{T}} \mathbf{H}_n[k] \rvert^2)}
	\end{equation}
	where $\rho$ denotes the SNR. As the steering angle gets closer to the relative angle between the RSU and the vehicle, the achievable rate increases. Since all subcarriers share the same beam eigenvectors  \cite{ofdmsubcarrier}, averaging achievable rates across all subcarriers as in \eqref{energy} will help increase the reliability in identifying the effective beams from $\mathbf{h}_n^{\rm{P}}$.

	\textbf{\textit{Step 2: Major beam detection by orthogonal matching pursuit (OMP)}}. The use of super-resolution DFT codebook can significantly improve the accuracy of the potential azimuth information of the main signal propagation paths. However, such information cannot be directly obtained by identifying the beam indices with top-$k$ highest energy from $\mathbf{h}_n^{\rm{P}}$. This is because of the beam energy dispersion jointly caused by the large number of beam energy generated by the super-resolution DFT codebook and the limited number of actual beams generated by the sub-6 GHz antenna array. To address this issue, we propose to apply the OMP over the $\mathbf{h}_n^{\rm{P}}$ to coarsely identify the major beams. The pseudo-code of this method is presented in \textbf{Algorithm 1}.
	\begin{figure}[!t]
		\label{alg}
		\renewcommand{\algorithmicrequire}{\textbf{Input:}}
		\renewcommand{\algorithmicensure}{\textbf{Output:}}
		\removelatexerror
		\begin{algorithm}[H]
			\caption{{\color{black}Major Beam Detection by OMP}}
			\begin{algorithmic}[1]
				{\color{black}
			    \REQUIRE Processed CSI matrix $\mathbf{h}_n^{\rm{P}}$, Original CSI matrix $\mathbf{H}_n$, Number of major beams to be detected $G$.   %%input
				\ENSURE Major beam index set $\mathcal{I}$  %%output
				\STATE Initialization: The residual $\mathbf{R}_n = \mathbf{H}_n$, beam energy vector $\hat{\mathbf{h}}_n^{\rm{P}}=\mathbf{h}_n^{\rm{P}}$, and $\mathcal{I}= \emptyset$. 
				\FOR {$j=1:G$}
				
				\STATE $m_j = \mathop{\arg\max}\limits_{i}\hat{\mathbf{h}}_n^{\rm{P}}[i], i \in [1, B]$
				\STATE $\mathcal{I}=\{\mathcal{I},m_j\}$
				\STATE $\hat{\mathbf{D}} = \mathbf{D}[:,{\mathcal{I}}]$, $\hat{\mathbf{H}}_n^{\rm{a}} = \mathbf{H}_n^{\rm{a}}[\mathcal{I},:]$
				\STATE $\mathbf{R}_n = \mathbf{R}_n  -  (\hat{\mathbf{D}}^{\rm{T}})^{\dagger}\hat{\mathbf{H}}_n^{\rm{a}}$
				\FOR {$q=1:B$}
				\STATE $\hat{\mathbf{h}}_n^{\rm{P}}[q]=\frac{1}{K_{\rm{LF}}}\sum_{k=1}^{K_{\rm{LF}}}\log_{2}{(1+ \rho \lvert \mathbf{d}_q^{\rm{T}} \mathbf{R}_n[k] \rvert ^2)}$
				\ENDFOR
				\ENDFOR
				\STATE Return the detected major beam index set $\mathcal{I}$.
			}
			\end{algorithmic}
		\end{algorithm}
	\end{figure}

	\textbf{\textit{Step 3: Exclusion of negligible beams}}. Note that only $k$ major beams  are detected in Step 2. To highlight the angular feature and boost the NN's learning capability, the corresponding elements in $\mathbf{h}_n^{\rm{P}}$ are retained and the other elements are set to $0$. We refer to this operation as \textit{exclusion of negligible beams}. The rationality of this operation will be demonstrated in the next subsection. The ultimate pre-processed CSI matrix $\mathbf{h}_n^{{\rm{P}}'} \in \mathbb{R}^{B\times 1}$ is expressed as
	\begin{equation}
		\label{energy2}
		\mathbf{h}_n^{{\rm{P}}'}[i]=\left\{
		\begin{array}{cl}
			\mathbf{h}_n^{\rm{P}}[i] & ,~i \in \mathcal{I} \\
			0  &  ,~otherwise
		\end{array} \right.
	\end{equation}

	\subsubsection{Angular Feature Extraction}
	
	 Following the above pre-processing, $\mathbf{h}_n^{{\rm{P}}'}$ is regarded as the wireless part of the multi-modal input data. To extract the angular feature from $\mathbf{h}_n^{{\rm{P}}'}$, the AFE module adopts the multilayer perceptron (MLP) architecture which contains $L_{\rm{A}}$ layers. Let $\bm{\nu}_n$ be the angular feature. $\bm{\nu}_n$ is obtained by
    \begin{equation}
    	\bm{\nu}_n  = \text{MLP}_{\rm A}(\mathbf{h}_n^{{\rm{P}}'},\mathbf{\Theta}_{\rm A}) = \varphi_{\rm A}^{L_{\rm A}}(
    	\varphi_{\rm A}^{L_{\rm A}-1}(...
    	\varphi_{\rm A}^{1}(\mathbf{h}_n^{{\rm{P}}'})...))
    \end{equation}
	where $MLP_{\rm{A}}$ represents the MLP network, $\mathbf{\Theta}_{\rm{A}}=\left\{\mathbf{W}_{\rm{A}},\mathbf{b}_{\rm{A}}\right\}$ represents its weights and bias, and $\varphi_{\rm{A}}^l(\cdot)$ represents the non-linear function of the $l$-th layer. $\varphi_{\rm{A}}^1$ can be expressed as
\begin{equation}	
	\label{NLFunction}
	\varphi_{\rm A}^{1}(\mathbf{h}_n^{{\rm{P}}'})  = f_\text{ReLU}( \mathbf{W}_{\rm{A}}^1\mathbf{h}_n^{{\rm{P}}'}+ \mathbf{b}_{\rm{A}}^1)
\end{equation}
	where $\mathbf{W}_{\rm{A}}^1$ and $\mathbf{b}_{\rm{A}}^1$ represent the weight and bias of the first layer. $f_{\rm{ReLU}}(\cdot)$ represents the ReLU function and can be expressed as $f_{\rm{ReLU}}(x) = \max(0,x)$.
	
	\begin{figure*}[!t]
		\centering
		\includegraphics[width=0.98\textwidth]{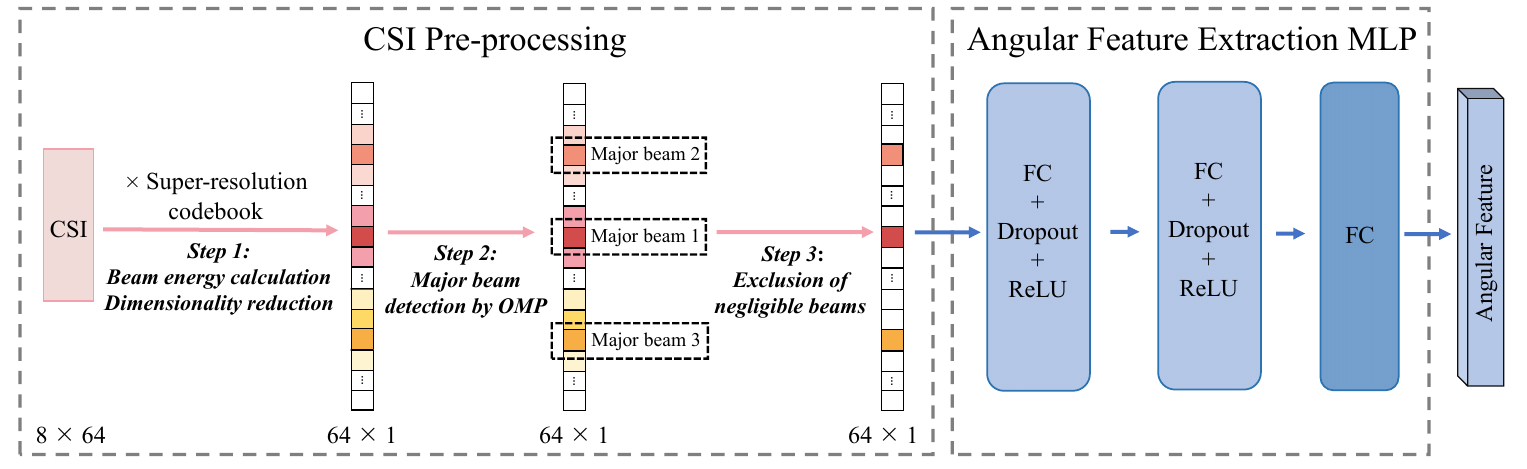}
		\caption{{\color{black}Processing flow for the AFE module.}}
		\label{csi}
	\end{figure*}
	
	For clarity, Fig. \ref{csi} shows the processing flow of the AFE module, where the CSI is first pre-processed and the angular feature is then extracted. As shown in Fig. \ref{csi}, the proposed AFE MLP consists of three hidden layers and one output layer. Let $L_{\rm{A}}$ be the number of neurons in the output layer which is also the size of the angular feature and can be adaptively adjusted.
	
	\subsubsection{Top-K Versus All}
\begin{figure}[!t]
		\centering
		\includegraphics[width=1\linewidth]{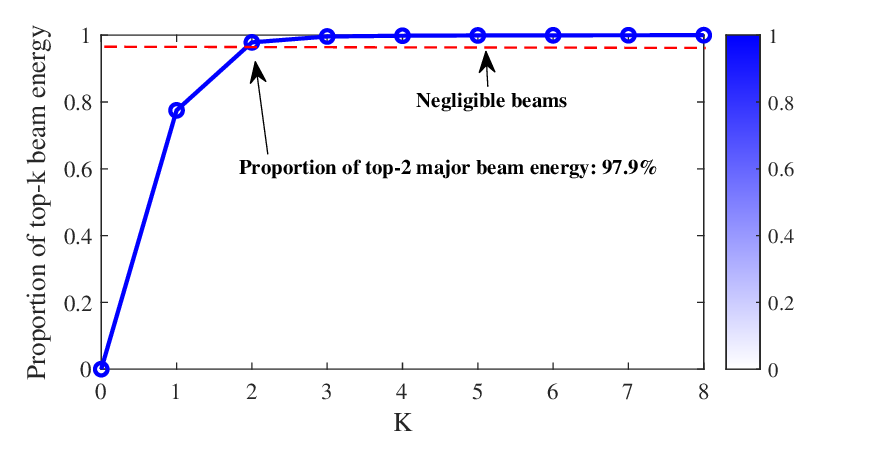}%
		\caption{Proportion of the top-$k$ major beam energy to the total energy.}
		\label{topk}
	\end{figure}

	The validity of prior CSI pre-processing is demonstrated via ``\textbf{\textit{Top-K Versus All}}" criterion.
	{\color{black}Specifically, the energy of the $i$-th beam (corresponding to the $i$-th steering vector) is given by $e_{i} = {\Vert (\mathbf{d}_{i}^{\rm{T}})^\dagger (\mathbf{D}^{\rm{T}}\mathbf{R}_n)[i] \Vert_{\rm{F}}^2}$. Then, the proportion of the sum of the top-$k$ major beam energy to the total energy is computed.} As shown in Fig.~\ref{topk}, the proportion dramatically increases as the number of selected major beams $k$ increases from $0$ to $2$. This can be attributed to the presence of a few main propagation paths between the RSU and the vehicle, with the total energy of the major beams whose directions are close to these paths accounting for the majority of the total energy. When $k$ becomes larger than $2$, the curve soon saturates, implying the power contributed by the remaining beams becomes negligible. As can be seen in Fig.~\ref{topk}, the top-$2$ beams account for $97.9\%$ of the total energy. Thus, they can be extracted as the latent features to streamline the processing with minimal performance loss. 
	
	\subsection{2D-Visual Feature Extraction Module}
	\label{VFE2}
    As mentioned in Section \ref{AFE}, CSI provides compressed electromagnetic environment information, and the angular feature potentially indicates the relative angle of the passing vehicle through LoS channel's azimuth.  However, the location information of the vehicle provided by angular feature is relatively rough, calling for a more explicit and adequate environmental awareness as a complement. Images can visually show the location and surrounding environment of the vehicle. The global visual features of the 2D environment provided by images can benefit the localization of the vehicle and the prediction of the future position. The 2D-visual features can supplement relevant information in a non-RF format, such as the locations of scatterers. To efficiently extract the 2D-visual feature, a 2D-visual feature extraction (VFE) module is incorporated into the MMFF-Net.
	\begin{figure*}[!t]
		\centering
		\includegraphics[width=1\linewidth]{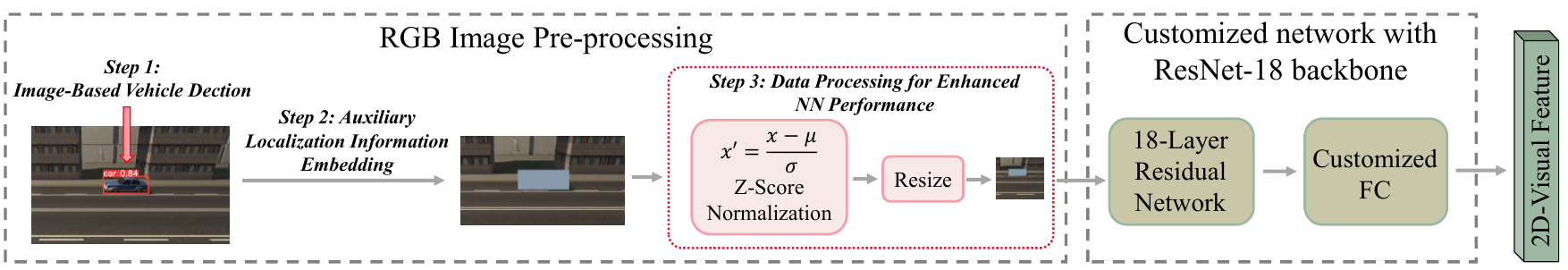}
		\caption{{\color{black}Processing flow for the VFE module.}}
		\label{VFE}
	\end{figure*}
	\subsubsection{Pre-processing of RGB Image}
	{\color{black}
	The proposed RGB image pre-processing method is dedicated to providing intuitive information that reflects the distribution of vehicle position relative to the RSU, which consists of three steps. Additionally, this method is designed to enable MMFF-Net to be directly applied to multi-user scenarios with minimal adjustments required, which will be elaborated in detail in Section~\ref{MLP}. 
	
	\textbf{\textit{Step 1: Image-Based Vehicle Detection.}} We employ the one-stage object detector YOLOv5 \cite{Yolo} to extract the bounding box coordinates of the vehicle detected in the image. Note that the pretrained YOLOv5 model has been fine-tuned on the ViWi dataset to accurately detect the vehicle in our study. Let $f_{\rm{YOLOv5}}(\cdot)$ represent the mapping performed by the detector. If $\mathbf{X}_n \in \mathbb{R}^{H \times W \times C}$ is the input image, where $H$, $W$ and $C$ represent the height, width and the number of color channels of the image, respectively, then, given $\mathbf{X}_n$ as the input, the output bounding box and objectness score of the vehicle can be expressed as:
	\begin{equation}
		\label{yolo}
		\{(u,v,h,\rho,s)\}=f_{\rm{YOLOv5}}(\mathbf{X}_n)
	\end{equation} 
	where $u$ and $v$ represent the coordinates of the corresponding bounding box, respectively; $h$ and $\rho$ represent the weight and aspect ratio of the detected bounding box, respectively; and $s$ represents the objectness score.
	
	\textbf{\textit{Step 2: Auxiliary Localization Information Embedding.}} In the context of motion tracking, it is essential to extract the foreground information from images, specifically the vehicle, to effectively guide the neural network in accurately predicting the vehicle’s future position. Also, the color information of the vehicle is redundant in this regard. To this end, we propose an auxiliary localization information embedding method based on the vehicle detection result to indicate the relative position of vehicle to the NN. This method involves converting the three color components of the pixels within the bounding box into the angle information of the vehicle's front, middle, and rear within the camera's field of view (FoV), denoted by $\gamma_{\rm{f}}$, $\gamma_{\rm{m}}$, and $\gamma_{\rm{r}}$. These parameters are calculated as follows: 
%	Based on the vehicle detection result, we further propose the auxiliary localization information embedding method since the foreground information of images, i.e., the vehicle, needs to be identified to better guide the NN to learn the vehicle's . Furthermore, the color information of the vehicle is redundant for BS regarding the motion tracking task. To this end, to indicate the relative position of vehicle to the network, this method converts three color components of the pixels within the bounding box into the angle information of the vehicle's front, middle, and rear within the camera's field of view (FoV), which is calculated by: 
	\begin{equation}
		\left\{
		\begin{array}{cl}
		\gamma_{\rm{f}} =  \gamma_{\rm{FoV}} \frac{[uW-\frac{h}{2}W]}{W}& \\
		\gamma_{\rm{r}} =  \gamma_{\rm{FoV}} \frac{[uW+\frac{h}{2}W]}{W}& \\
			\gamma_{\rm{m}} = \frac{\gamma_{\rm{f}} + \gamma_{\rm{r}} }{2}&
		\end{array} 
		\right.
	\end{equation}
   where $\gamma_{\rm{FoV}}$ is the camera's FoV and $W$ is width of the image, respectively. Then, $(255\frac{\gamma_{\rm{f}}}{\gamma_{\rm{FoV}}},255\frac{\gamma_{\rm{m}}}{\gamma_{\rm{FoV}}},255\frac{\gamma_{\rm{r}}}{\gamma_{\rm{FoV}}})$ is used as the RGB channel values for the pixels in the bounding box, as shown in Fig.~\ref{VFE}.
    
	\textbf{\textit{Step 3: Data Processing for Enhanced NN Performance.}} Finally, to enhance network performance and ensure a stable training process, the values of features in the processed RGB image are appropriately normalized to achieve a similar range.} Specifically, the Z-score normalization is adopted to normalize the three channels of the processed RGB images as
	\begin{equation}
		\label{normal}
		\hat{\mathbf{X}}_n^{(j)}=\frac{\mathbf{X}_n^{(j)}-\mu_n^{(j)}}{\sigma_n^{(j)}}, j=1,2,3
	\end{equation} 
	where $\hat{\mathbf{X}}_n^{(j)}$ is the $j$-th normalized channel, $\mu_n^{(j)}$ and $\sigma_n^{(j)}$ are the mean value and the standard deviation of the pixel value in the $j$-th channel, respectively. Then, the normalized RGB images are resized to $224 \times 224$\footnote{$224 \times 224$ is one of the commonly adopted sizes in computer vision field.} to reduce unnecessary computation for NN since there is plenty of redundant information in the original images. 
	
	\subsubsection{2D-Visual Feature Extraction}
    After the RGB image pre-processing, $\hat{\mathbf{X}}_n^{\rm{P}}$ is regarded as the visual part of the multi-modal input data. In the computer vision field, many powerful NNs have been designed for image feature extraction and analysis \cite{Ngiam2011}, serving tasks like target detection and tracking (e.g., Res-Net \cite{he2016deep} and AlexNet \cite{krizhevsky2017imagenet}). To efficiently extract the deep feature of RGB images, the ResNet-18 \cite{he2016deep} is adopted in VFE module for its capability of extracting more abundant features while simplifying the learning difficulty and preventing the vanishing gradient issue. Note that the adopted ResNet-18 is customized to adaptively adjust the size of the extracted 2D-visual feature in the MMFF-Net. Specifically, its output {\color{black}fully connected (FC)} layer is replaced by one with customized length $L_{\rm{V}}$ which also represents the length of the 2D-visual feature. For clarity, the processing flow for VFE module is shown in Fig.~\ref{VFE}.

	\subsection{Distance Feature Extraction Module}
	\label{DFE2}
	
	Distance information is critical for accurate position prediction, but the AFE and VFE modules  do not provide precise distance information. This lack of information limits the ability to reconstruct the 3D environment and predict the vehicle's exact position. To address this issue, we propose incorporating radar ranging results as an essential sensing modality that naturally complements the semantic information of RGB images by providing 3D geometric shapes for 2D-visual information. To address the lack of radar data in the ViWi dataset, we simulate radar sensing by adding Gaussian noise to depth maps. While the amount of distance data in simulation exceeds that of actual radar ranging, our goal is to showcase the benefits of incorporating distance information in position prediction. We will describe the pre-processing of depth maps and the extraction of distance features in the distance feature extraction (DFE) module.
	
	\subsubsection{Pre-processing of Depth Map}
	At present, the processing methods of depth map is generally regarding it as an additional channel of the RGB image. This pre-processing approach is not adopted in DFE module. The reasons are as follows. First, the commonly adopted approach assumes that the distance data is accurate and perfectly aligned with RGB image pixels, which is difficult to achieve by practical radar sensing. Second, distance data in depth maps are used to mimic frequent radar ranging results in this scheme. The multi-modal feature extraction and fusion method needs to be universal to radars of various operating frequencies in practical applications. To this end, the depth map is treated as an independent data source. 
                  	
	To boost the learning performance of the MMFF-Net, the order of magnitude of input multi-modal data needs to be approximately the same. Also, the values of features within the depth map need to be aligned comparably. To this end, the min-max normalization is adopted to normalize the input depth map. Let $\mathbf{Y}_n$, $\hat{\mathbf{Y}}^{\rm{P}}_n$, and $\mathbf{Z}$ be the raw, the pre-processed depth map, and Gaussian noise with zero mean and variance $N_0$, respectively. $N_0$ is set as $0.1$ in this paper. Then, the $\hat{\mathbf{Y}}^{\rm{P}}_n$ is obtained as
	\begin{equation}
		\label{minmax}
		\hat{\mathbf{Y}}^{\rm{P}}_n=\frac{(\mathbf{Y}_n+\mathbf{Z})-\min(\mathbf{Y}_n+\mathbf{Z})}{\max(\mathbf{Y}_n+\mathbf{Z})-\min(\mathbf{Y}_n+\mathbf{Z})}
	\end{equation}

	\subsubsection{Distance Feature Extraction}
	The radar ranging results are spatially-independent discrete values. In order to make the frequent radar ranging results reflect the overall spatial structure of the environment, the radar ranging results are arranged in the form of images according to the ranging direction. In this scheme, the depth map is equivalent to the radar ranging results with spatial correlation.
	
\begin{figure}[!t]
		\centering
		\includegraphics[width=1\linewidth]{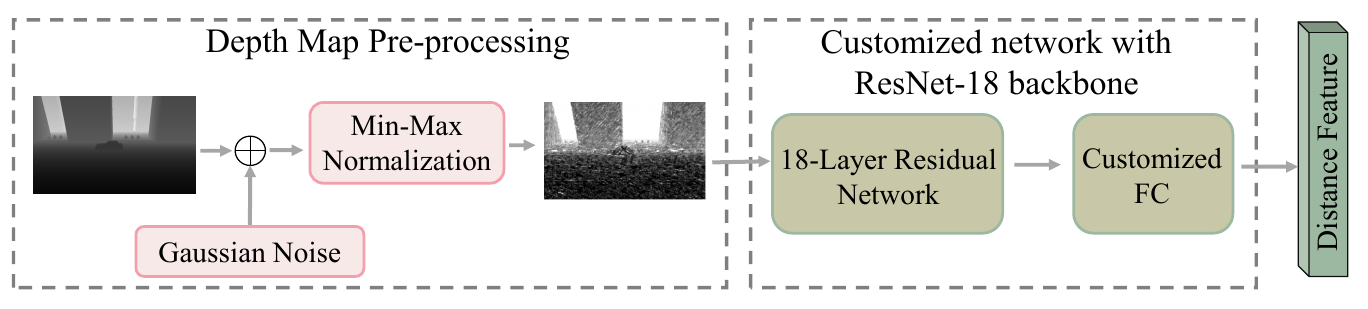}
		\caption{Processing flow for the DFE module.}
		\label{DFE}
	
	\end{figure}

	Similar to the VFE module, the ResNet-18 is adopted and customized in DFE module to adaptively adjust the size of the extracted distance feature. The output FC layer of ResNet-18 is replaced by one with length $L_{\rm{D}}$ which also represents the length of distance feature. For clarity, the processing flow for DFE module is shown in Fig. \ref{DFE}.
	
	\section{Multi-Modal Feature Fusion and Beamforming Angle Prediction}
	
	\label{IV}

	In this section, we will introduce the adaptive weighted fusion method for combining the angular feature, 2D-visual feature, and distance feature. Following that, we will elaborate on how to proactively predict the beamforming angle based on the two regression models in the MMFF-Net.

    \subsection{Adaptive Weighted Fusion of Multi-Modal Feature}

	%简单介绍 并行特征的拼接
	The multi-modal data used in this scheme contains completely different semantic information. The 2D-visual feature, distance feature, and angular feature need to be preserved entirely to prevent the mixing of different features. Therefore, a tensor concatenation operation is first applied to combine the three features and create an informative representation of the multi-modal feature.
	
	Let ${\rm{AFE}(\cdot)}$, ${\rm{VFE}(\cdot)}$, and ${\rm{DFE}(\cdot)}$ represent the feature extraction processing of the proposed AFE module, VFE module, and DFE module, respectively. Let $\bm{\nu}_n$, $\mathbf{\chi}_n$, $\bm{\tau}_n$, and $\mathbf{f}_n$ represent the angular feature, 2D-visual feature, distance feature, and multi-modal feature, respectively. Then, the multi-modal feature $\mathbf{f}_n$ can be obtained through \eqref{ang}~-~\eqref{fusion}
	\begin{subequations}
		\begin{align}
			
			& \bm{\nu}_n = {\rm{AFE}}(\mathbf{h}_n^{\rm{P}'}) \label{ang} \\

			& \bm{\chi}_n = {\rm{VFE}}(\hat{\mathbf{X}}^{\rm{P}}_n) \label{image}\\

			& \bm{\tau}_n = {\rm{DFE}}(\hat{\mathbf{Y}}^{\rm{P}}_n) \label{dp} \\

			& \mathbf{f}_n = \bm{\nu}_n \oplus \bm{\chi}_n \oplus \bm{\tau}_n \label{fusion} 
		\end{align}
	\end{subequations}
	
	However, treating different features as equally important may potentially lead to performance loss. Therefore, we introduce an adaptive weight learning network (AWLN) module in the MMFF-Net to learn and assign appropriate weights to the multi-modal features. Let $\varphi_{\rm{w}}^l(\cdot)$ be the non-linear function of the $l$-th layer in AWLN module with ReLU serving as the activation function. $\varphi_{\rm{w}}^l(\cdot)$ is similarly defined as in \eqref{NLFunction}. Then the weight vector $\mathbf{w}_n$ output by AWLN can be expressed as
	\begin{equation}
	\label{att}
    \mathbf{w}_n  = f_{\rm{Sig}}(\mathbf{W}_{\rm{L}}(\varphi_{\rm{w}}^2(\varphi_{\rm{w}}^1(\mathbf{f}_n)))+\mathbf{b}_{\rm{L}})
   \end{equation}
    where $\mathbf{W}_{\rm{L}}$ and $\mathbf{b}_{\rm{L}}$ are the weight and bias of the last FC layer in the AWLN module; $f_{\rm{Sig}}(\cdot)$ is the Sigmoid function and can be expressed as $f_{\rm{Sig}}(x) = \frac{1}{1+e^{-x}}$. $f_{\rm{Sig}}(\cdot)$ is served as the activation function of the last FC layer to ensure that the values in the weight vector are between $0$ and $1$. 
    
    In fact, the intra-modality weight difference can be omitted and we only need to focus on the inter-modality weight difference. Therefore, we calculate the mean values of the angular feature weight, 2D-visual feature weight, and distance feature weight from the weight vector $\mathbf{w}_n$, denoted by $w_{\rm{A}}$, $w_{\rm{V}}$, and $w_{\rm{D}}$, respectively. They are calculated by $w_{\rm{A}} = \frac{1}{L_{\rm{A}}}  \sum_{i=1}^{L_{\rm{A}}}\mathbf{w}_n(i)$, $w_{\rm{V}} = \frac{1}{L_{\rm{V}}}  \sum_{i=1+L_{\rm{A}}}^{L_{\rm{A}}+L_{\rm{V}}}\mathbf{w}_n(i)$, and $w_{\rm{D}} = \frac{1}{L_{\rm{D}}} \sum_{i=1+L_{\rm{A}}+L_{\rm{V}}}^{L_{\rm{A}}+L_{\rm{V}}+L_{\rm{D}}} \mathbf{w}_n(i)$. Then, we adopt a modality-wise product\footnote{{\color{black}Akin to element-wise product, the term ``modality-wise product" hereby refers to the operation of multiplying all the feature values corresponding to a certain modality by a certain value.}} to obtain the weighted multi-modal feature $\mathbf{f}_n^{\rm{w}}$:
\begin{equation}
	\label{weigthed feature}
	\mathbf{f}_n^{\rm{w}}  = w_{\rm{A}}\bm{\nu}_n \oplus w_{\rm{V}}\bm{\chi}_n \oplus w_{\rm{D}}\bm{\tau}_n
\end{equation}
	
\begin{figure}[!t]
		\centering
		\includegraphics[width=1\linewidth]{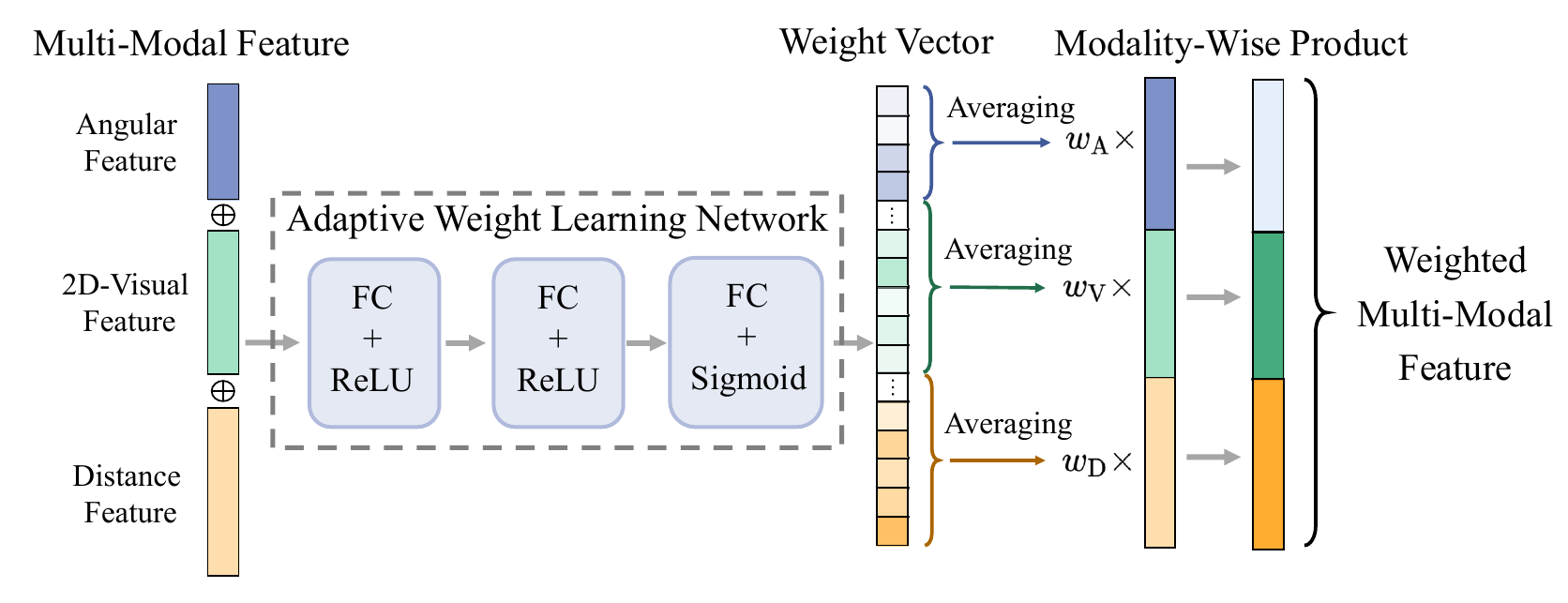}
		\caption{{\color{black}Processing flow for the multi-modal feature fusion module.}}
		\label{wf}
	\end{figure}

	For clarity, the processing flow for the AWLN module is depicted in Fig. \ref{wf}. The weighted multi-modal features $\mathbf{f}_n^{\rm{w}}$ and $\mathbf{f}_{n-1}^{\rm{w}}$ are then input into the X-axis prediction MLP and Y-axis recurrent prediction module to predict the beamforming angle, as elaborated in the next subsection.	

	\subsection{Beamforming Angle Prediction}
	%介绍x轴（变道轴）的预测
    In the studied scenario, the vehicle is allowed to drift laterally rather than moving in an ideal straight lane. The vehicle's position tracking is equivalent to the 2D coordinate $(\hat{x}_{{n}+1},\hat{y}_{{n}+1})$ prediction of the vehicle. Assume that the vehicle keeps moving in a straight lane on the $x$-axis, and drifts laterally with a certain probability on the $y$-axis.
\subsubsection{Temporal Feature Extraction and Recurrent Prediction for Y-Coordinates Prediction}
    When the KF-based and the EKF-based state tracking schemes are applied to irregular motion scenarios, the tracking results when the motion state changes have large errors. The motion parameters of vehicles naturally exhibit correlation at adjacent time slots and the correlation tends to decrease with the increase of time interval. Consequently, we propose to extract the temporal feature of the vehicle's movement and conduct recurrent prediction of the vehicle's $y$-coordinates.  By learning the short-term features of the vehicle's movement from the historical multi-modal data, there will be no large error in the coordinate prediction results when the vehicle drifts laterally.  	
    
    Long short-term memory and gated recurrent unit (GRU) networks were designed to solve the problem of vanishing gradient in recurrent neural networks for time series prediction \cite{DL}. In the MMFF-Net, GRU is adopted to extract the temporal feature of the vehicle's movement and conduct the recurrent prediction. Assume that each GRU layer contains $T_{\rm{u}}$ GRU units.The input to the GRU layers in the MMFF-Net consists of sequences of weighted multi-modal features, denoted as  $\mathbf{f}^{\rm{w}}_{1,2,\cdots,T_{\rm{u}}}$. Then, the output of the GRU layers can be expressed as
    \begin{equation}
    	\label{GRU}
    		\bm{O}_{(T_{\rm{u}})} = F_{\rm{GRU}}(\mathbf{f}_{\rm{1}}^{\rm{w}},\mathbf{f}_{\rm{2}}^{\rm{w}},\cdots,\mathbf{f}_{(T_{\rm{u}})}^{\rm{w}})
    \end{equation}

    GRU utilizes the update gate and reset gate to control the flow of information and determines the update of the hidden state. Given the page limit, the relationships between the input sequences $\mathbf{f}_{1,2,\cdots,T_{\rm{u}}}^{\rm{w}}$ and output $\bm{O}_{(T_{\rm{u}})}$ are not detailed in this paper. We only give the relationship between $\bm{O}_{(T_{\rm{u}})}$, which is also the output of the last unit, and other information as follows
%    In the $t$-th unit of a certain layer ($1\leq t \leq T_{\rm{u}}$), the relationships between the input $\mathbf{x}^{(t)}$ and output $\mathbf{y}^{(t)}$ can be expressed through \eqref{GRU1} - \eqref{GRU4}. Let $\mathbf{z}^{(t)}$ and $\mathbf{r}^{(t)}$ be the values of the $t$-th update gate and reset gate in GRU. $f_{\rm{Sig}}$ and $f_{\rm{Tanh}}$ represent the Sigmoid and Tanh function. Then, \eqref{GRU1} and \eqref{GRU2} are used to obtain $\mathbf{z}^{(t)}$ and $\mathbf{r}^{(t)}$.
%    \begin{equation}
%    	\label{GRU1}
%    	\mathbf{z}^{(t)}= f_{\rm{Sig}}(\mathbf{w}_{\rm{z}}\mathbf{x}^{(t)}+\mathbf{u}_{\rm{z}}\mathbf{y}^{(t-1)})
%    \end{equation}
%    \begin{equation}
%    	\label{GRU2}
%    	\mathbf{r}^{(t)}= f_{\rm{Sig}}(\mathbf{w}_{\rm{r}}\mathbf{x}^{(t)}+\mathbf{u}_{\rm{r}}\mathbf{y}^{(t-1)})
%    \end{equation}
%    where $\mathbf{w}_{\rm{z}}$, $\mathbf{w}_{\rm{r}}$, $\mathbf{u}_{\rm{z}}$ and $\mathbf{u}_{\rm{r}}$ are the weights and biases of the update gate and reset gate, respectively.
%    
%    The information retaine nd by the $t$-th unit is calculated by \eqref{GRU3}
%    \begin{equation}
%    	\label{GRU3}
%    	\widetilde{\mathbf{o}}^{(t)}= f_{\rm{Tanh}}(\mathbf{w}_{\rm{o}}\mathbf{x}^{(t)}+\mathbf{u}_{\rm{o}}(\mathbf{r}^{(t)}\otimes\mathbf{y}^{(t-1)}))
%    \end{equation}
%    where $\mathbf{w}_{\rm{o}}$ and $\mathbf{u}_{\rm{o}}$ represent the linear transformations performed on $\mathbf{x}^{(t)}$ and $\mathbf{y}^{(t-1)}$, respectively.
    \begin{equation}
    	\label{GRU4}
    	\bm{O}_{(T_{\rm{u}})}=(1-\bm{z}_{(T_{\rm{u}})})\otimes\bm{O}_{(T_{\rm{u}}-1)}+\bm{z}_{(T_{\rm{u}})}\otimes\widetilde{\bm{O}}_{(T_{\rm{u}})}
    \end{equation}
    where $\bm{z}_{(T_{\rm{u}})}$ is the value of the $T_{\rm{u}}$-th update gate in GRU and $\widetilde{\bm{O}}_{(T_{\rm{u}})}$ is the information retained by the $T_{\rm{u}}$-th unit.

%    In the MMFF-Net, the multi-modal data collected at the current time instance and the previous one are used to construct time series so that GRU can learn the motion feature of the vehicle. They are first input into the multi-modal feature extraction and fusion module, where the 2D-visual, distance, and angular features are extracted and fused. Then, the multi-modal feature at the current time instance is input into the X-axis prediction MLP to predict the $x$-coordinate. The multi-modal features from the current time instance and the previous one are stacked and input into Y-axis recurrent prediction module to predict the $y$-coordinate. 
    
    \subsubsection{MLP Prediction for X-Coordinates Prediction}
    \label{MLP}
    The prediction of the vehicle's $x$-coordinates is carried out by a four-layer MLP rather than a recurrent NN, since the feature of the vehicle's movement on the $x$-axis is easily learned by NN. This design ensures prediction accuracy while reducing the computational load and accelerating the convergence of NN.

\begin{figure*}[!t]
	\centering
	\includegraphics[width=0.96\textwidth]{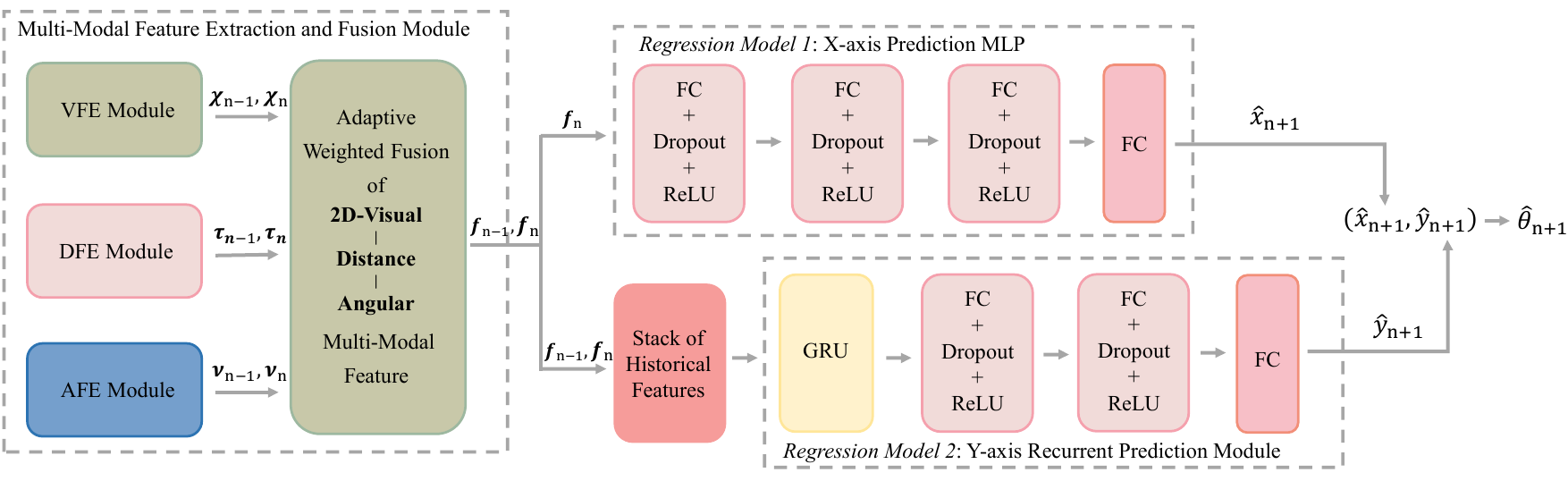}
	\caption{{\color{black}The proposed MMFF-Net is composed of three main modules: (i) {multi-modal feature extraction and fusion module}; (ii) {Y-axis recurrent prediction module}; and (iii) X-axis prediction MLP.}}
%	(i) \textbf{multi-modal feature extraction and fusion module}, for extracting and combining the 2D-visual, distance, and angular features from the multi-modal data; (ii) \textbf{Y-axis recurrent prediction module}, for predicting the $y$-axis coordinate based on the stacked historical multi-modal features; and (iii) \textbf{X-axis prediction MLP}, for utilizing the multi-modal feature at the current time slot to predict the $x$-axis coordinate.
	\label{scheme}
    \end{figure*}

    With the predicted vehicle's 2D coordinate, the beamforming angle in the next time instance $\hat{\theta}_{n+1}$ can be determined by $\hat{\theta}_{n+1} = \arctan(\frac{\hat{x}_{{n}+1}-x_{RSU}}{\hat{y}_{{n}+1}})$ with $x_{RSU}$ being the x-coordinate of the RSU. For clarity, Fig. \ref{scheme} shows the overall block diagram of the MMFF-Net.

    \textit{Remark 1}: 
    The proposed MMFF-Net can be extended to multi-user scenarios. By allocating orthogonal subcarriers to different vehicles, the system can differentiate between them and effective angular features can be extracted from CSI matrices corresponding to different subcarriers. {\color{black}The VFE and DFE modules can initially employ multi-object detection by the RSU to identify multiple vehicles within the FoV. Assuming that the user association has already been performed using existing schemes \cite{MU-matching, differentiation_of_multi_v, user_identification}, the proposed RGB image and depth map processing method can then be applied by focusing on the object detection result and corresponding ranging information associated with a specific vehicle, while considering other vehicles as background elements. Through this approach, MMFF-Net enables simultaneous tracking of multiple vehicles and accomplishes proactive beamforming in multi-user scenarios.}
    
%    BS can achieve the differentiation of multiple vehicles by performing the following operations \textit{in the initial block} . BS first conducts multi-vehicle detection and matches the bounding boxes with the depth values or radar ranging results to estimate the relative angle of departure (AoD) of each vehicle. Meanwhile, the vehicles also estimate the AoDs based on downlink training and feed them back to BS through uplink. BS then matches and associates the feedback AoDs with the estimated ones to distinguish multiple vehicles it serves and continuously track them.  
    
%	\vspace{-0.1cm}
\section{Experimental Setup}
\label{V}
This section introduces the evaluation dataset and metrics, as well as the network architecture and training methodology for MMFF-Net.

\subsection{Dataset Overview}
\label{Dataset}	
\begin{figure}[!t]
	\centering
	\includegraphics[height=4cm,width=7.3cm]{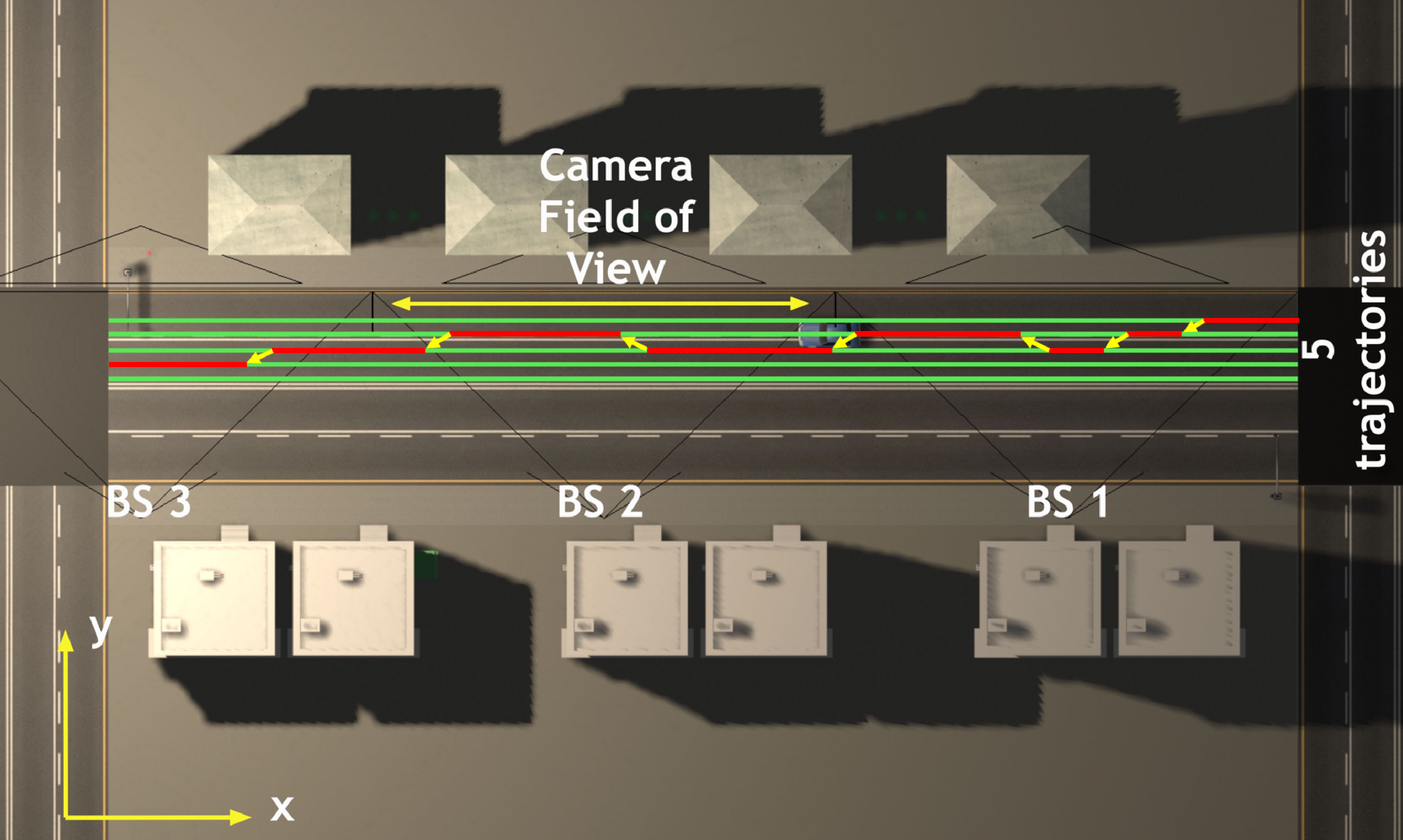}
	\caption{An example of the vehicle's trajectory where the vehicle randomly drifts. This example is constructed based on the ``dist\_cam" scenario.}
	\label{viwi3}
    \end{figure}

The dataset used for testing is based on the ViWi data-generation framework \cite{alrabeiah2020viwi}. The ``dist\_cam" scenario acts as the basis, where a single car drives through a city street. The FoV of the camera in this scenario is $100^{\circ}$. The hyper-parameters for channel generation are presented in Table~\ref{communication setup}. {\color{black}We simulate the sub-6 GHz CSI based on the wireless channel parameters provided by ViWi dataset. To emulate the rich multi-path components in sub-6 GHz band, we reset the path gains while keeping the other parameters of paths unchanged such as AoA and time of arrival. The parameters of the LoS component remain unchanged, assuming that there are two paths with a $2-5$ dB random gain reduction compared to the LoS component and three paths with a $5-10$ dB random gain reduction. The rationality of this methodology is supported by a comprehensive channel measurement campaign conducted in diverse scenarios \cite{mmMAGIC}, which indicated that the geometries of main propagation paths of two frequency bands are almost similar.} Furthermore, to make the dataset more realistic, the vehicle is assumed to randomly move to adjacent lanes during its movement. The complete trajectory remains continuous except for some time instances of lane changing. This randomness can also be regarded as the vehicle's drifting. Fig.~\ref{viwi3} displays an example of the vehicle’s trajectory where it randomly drifts. {\color{black}For the construction of training and testing datasets, we randomly generate two entirely different trajectories, where the vehicle's drifting behaviors occur randomly and differently.}

%   \vspace{-0.3em}
	\begin{table}[!htp]
		\setlength{\abovecaptionskip}{0.1cm} 
		\renewcommand\arraystretch{0.3} 
		\centering
		\caption{Hyper-parameters for channel generation}
		\label{communication setup}
		\begin{tabular}{c|c|c}
			\toprule[0.35mm]
			\textbf{Parameter}	& {\textbf{mmWave Band}}  & {\textbf{Sub-6 GHz Band}}  \\
			\midrule[0.2mm]
			Scenario name	& \multicolumn{2}{c}{dist\_cam} \\ 
			\midrule[0.2mm]
			Active BSs & \multicolumn{2}{c}{1} \\ 
			\midrule[0.2mm]
			{\color{black} \makecell[c]{Codebook size $B$} }  & -- & 64 \\
			 
			\midrule[0.2mm]			
			Transmit antennas (x, y, z)	& (8/16/32,1,1)  & (8,1,1) \\  
			\midrule[0.2mm]			
			Receive antennas (x, y, z)	& (8/16/32,1,1) & (1,1,1) \\
	
			\midrule[0.2mm]			
            Bandwidth (GHz)	&  -- & 0.08 \\
				
			\midrule[0.2mm]		
			Antenna spacing	&  \multicolumn{2}{c}{0.5} \\ 
			\midrule[0.2mm]			
			OFDM sub-carriers 	& -- & 64  \\		
			\midrule[0.2mm]			
			OFDM limit	&   -- &  32 \\
			\midrule[0.2mm]			
	        Paths	&   -- & 15 \\
			\bottomrule[0.35mm]
		\end{tabular}	
	\end{table}

%	\vspace{-0.5em}
	\subsection{Signal Model}
	\label{Signal model}
	At the $n$-th epoch, RSU sends a downlink directional data stream $s_n(t)$ to the vehicle using beamforming matrix $\mathbf{f}_n\in \mathbb{C}^{N^{\rm{HF}}_{\rm{t}}\times 1}$ (equivalent to the transmit beamformer for the single vehicle). The beamforming matrix $\mathbf{f}_n$ is designed based on the intended direction. Assume that the beamforming direction $\hat{\theta}_n$ is already obtained by the proposed scheme, the beamforming matrix is given by $\textbf{f}_n=\mathbf{a}(\hat{\theta}_n)$, with $\mathbf{a}(\theta)=\frac{1}{\sqrt{N^{\rm{HF}}_{\rm{t}}}}[1,e^{-{\rm{j}}\pi\cos \theta},\cdots,e^{-{\rm{j}}\pi(N^{\rm{HF}}_{\rm{t}}-1)\cos \theta}]^{\rm{T}}$ representing the transmit steering vector of RSU's antenna array, which is assumed to adopt half-wavelength antenna spacing. Likewise, let $\mathbf{b}(\theta)=\frac{1}{\sqrt{M^{\rm{HF}}_{\rm{r}}}}[1,e^{-{\rm{j}}\pi\cos \theta},\cdots,e^{-{\rm{j}}\pi(M^{\rm{HF}}_{\rm{r}}-1)\cos \theta}]^{\rm{T}}$ be the receive steering vector.

	At the $n$-th epoch, the vehicle forms a receive beamformer $\mathbf{w}_n$ according to the predicted angle to receive the signals transmitted by RSU. The receive signal is expressed as \cite{gao2020estimating}
	\begin{equation}
		\begin{aligned}
			\label{receive signal}
			c_n(t)= \tilde\kappa\sqrt{p_n}\alpha_n e^{{\rm{j}}2\pi\mu_nt}  \mathbf{w}_n^{\rm{H}}\mathbf{b}(\theta_n)\mathbf{a}^{\rm{H}}(\theta_n)\mathbf{f}_n s_n(t) +z_{\rm{c}}(t)
		\end{aligned}
	\end{equation}
	where $p_n$ is the transmit power; $\tilde\kappa=\sqrt{N_{\rm{t}}^{\rm{HF}} M_{\rm{r}}^{\rm{HF}}}$ is the antenna array gain; $\alpha_n$ is the reflection coefficient, $z_{\rm{c}}(t)$ is the Gaussian noise term which has zero mean and variance $\sigma_{\rm{c}}^2$. $\textbf{w}_n=\mathbf{b}(\hat{\theta}_n)$ is the receive beamformer that the vehicle prepares according to the predicted relative angle between the vehicle and RSU at the $n$-th epoch. $\mu_n$ denotes the Doppler frequency and is affected by the vehicle's velocity $v_n$, relative angle with RSU $\theta_n$, and carrier frequency $f_{\rm{c}}$, i.e., $\mu_n=\frac{v_n\cos\theta_nf_{\rm{c}}}{c}$. $\alpha_n =\tilde\alpha d_n^{-1} e^{{\rm{j}} \frac{2\pi f_{\rm{c}}}{c}d_n} $ denotes the LoS channel coefficient. $\tilde\alpha$ representing the reference power gain factor which is assumed to be known to RSU by calculating the channel power gain at the reference distance and $\frac{2\pi f_{\rm{c}}}{c}d_n$ represents the phase of the LoS channel. 
	
	Suppose that the transmit signal has a unit power, then the SNR of the signal received by the vehicle is given by ${\rm SNR}_n = \frac{p_n{\lvert\tilde\kappa\alpha_n\mathbf{b}(\hat{\theta}_n)^{\rm{H}}\mathbf{b}(\theta_n)\mathbf{a}^{\rm{H}}(\theta_n) \mathbf{a}(\hat{\theta}_n) \rvert}^2}{\sigma^2_{\rm{c}}}$. The achievable rate of the established link is given by $R_n = \log_2(1+{\rm SNR}_n)$. As discussed above, the achievable rate depends on the transmit and receive beamformers, which are designed based on the predicted relative angle between the vehicle and RSU. The achievable rate is maximized if the predicted beamforming angle is perfectly consistent with the actual angle, i.e., $\theta_n=\hat{\theta}_n$, yielding a SNR upper bound as ${\rm SNR}_{\rm{max}} = \frac{p_n{\lvert\tilde\kappa\alpha_n \rvert}^2}{\sigma^2_{\rm{c}}}$.

	\subsection{Network Configuration}
	\label{networkpara}
	
	Table \ref{DL setup} shows the hyper-parameters used for designing the network, including the sizes of multi-modal features and the architecture details of multiple NN modules in the MMFF-Net. {\color{black}The size of angular feature is set smaller than the others since the vehicle's position information it contains may not be as intuitive and reliable as that present in visual and distance features, especially in complex environments. In terms of NN architectures, we evaluated NNs with various depths and widths, and then determined the architecture used in this paper based on their performance and computational complexity.}  The AFE MLP has two hidden layers and one output layer, with all dropout layers having a deactivation rate of $0$. The X-axis prediction MLP has three hidden layers and one output layer, with all dropout layers having a deactivation rate of $p=0$. The Y-axis recurrent prediction module has two layers of GRUs, each with $r$ GRU units. The MLP connected to the second GRU layer has two hidden layers and one output layer, with two dropout layers having a deactivation rate of $p=0.1$.
	 %The ReLU function serves as the activation function in the MMFF-Net.
	Table \ref{DL setup2} lists the hyper-parameters used for fine-tuning the model. The training of MMFF-Net is performed by PyTorch, using the adaptive moment estimation (ADAM) optimizer \cite{Adam}. 
%	In practical applications, MMFF-Net is first trained offline and then deployed at RSU with abundant computing resources. Consequently, the online proactive beamforming only requires simple calculations on low-dimensional pre-processed input data and does not bring high computation complexity.
%		   \vspace{-0.1em}
		\begin{table}[!htp]
		\setlength{\abovecaptionskip}{0.1cm} 
		\renewcommand\arraystretch{0.9} 
		\centering
		\caption{Hyper-parameters for network design}
		\label{DL setup}
		\resizebox{\linewidth}{!}{
			\begin{tabular}{c|c}
				\toprule[0.35mm]
				\textbf{Parameter}	&\textbf{Value}  \\
				\midrule[0.2mm]
				The size of multi-modal features@[$L_{\rm{A}}$, $L_{\rm{V}}$, $L_{\rm{D}}$]	& [16, 256, 256] \\ 
				\midrule[0.2mm]
				GRU units per layer $r$	& 16 \\ 
				\midrule[0.2mm]
				\makecell[c]{Neurons in hidden layers\\of AFE MLP@[layer1, layer2, layer3]}	& [64, 32, 16] \\ 
				\midrule[0.2mm]
				\makecell[c]{Neurons in hidden layers of X-axis \\prediction MLP@[layer1, layer2, layer3, layer4]} &  [528, 256, 128, 64] \\ 
				\midrule[0.2mm]			
				\makecell[c]{Neurons in hidden layers of Y-axis recurrent\\ prediction module@[layer1, layer2, layer3]}	& [16, 32, 16] \\ 	
				\midrule[0.2mm]	
				{\color{black}	
				\makecell[c]{Neurons in hidden layers of  \\ AWLN module@[layer1, layer2, layer3]}}	& 	{\color{black}[528, 264, 528]} \\ 
			
				\bottomrule[0.35mm]
			\end{tabular}	
		}
	\end{table}
%    \vspace{-0.1em}
	\begin{table}[!htp]
		\setlength{\abovecaptionskip}{0.1cm} 
		\renewcommand\arraystretch{0.8} 
		\centering
		\caption{Hyper-parameters for network fine-tuning}
		\label{DL setup2}
		\begin{tabular}{c|c}
			\toprule[0.35mm]
			\textbf{Parameter}	&\textbf{Value}  \\
			\midrule[0.2mm]
			Batch size	& [1, 5, 10] \\ 
			\midrule[0.2mm]
			Learning rate & $3 \times 10^{-4}$ \\ 
			\midrule[0.2mm]			
			Learning rate scheduler	& Epochs 15 and 30 \\ 	
			\midrule[0.2mm]			
			Learning-rate decaying factor & 0.3 \\
			\midrule[0.2mm]		
			Epochs  & 50 \\ 	
			\midrule[0.2mm]		
			Optimizer  & ADAM \\ 
			\midrule[0.2mm]			
			Loss funtion  & SmoothL1Loss \\ 	
			\bottomrule[0.35mm]		
		\end{tabular}	
	\end{table}

%	\vspace{-0.2cm}
	\subsection{Benchmarks}
	\label{Baseline}
	In this paper, extensive simulation results are provided and comprehensive comparisons between the proposed scheme with several existing benchmarks are conducted.\footnote{{\color{black}Regarding the benchmark schemes, the experimental parameter settings are set to be consistent with those given in \cite{shaham2019fast,liu2020radar,mu2021integrated,alrabeiah2020millimeter}  and the measurements in KF-T are set to ground truth values in the following experiments to ensure fair comparisons.}} In order to highlight the superiority and importance of multimodality fusion, three uni-modal data based schemes are designed and fine-tuned to perform the same proactive beamforming task.

	\begin{itemize}	
	\item \textbf{KF-based tracking (KF-T)}: A typical KF is adopted to track the vehicle's motion parameters, where ground truth values added with noises serve as the observation values.
	
	\item \textbf{EKF-based tracking (EKF-T)} \cite{shaham2019fast}:  In \cite{shaham2019fast}, an EKF-based method is proposed where the observation values are the transmit signal echoes. The state variables are the vehicle's position and velocity, as well as the channel coefficient. 
%	Note that the commonly used state variables, angles of arrival and angles of departure, are not adopted in this scheme since they lead to high complexity in the calculation of Jacobians. 
	The vehicle is assumed to move in an ideal straight line in the considered scenario.
	
	\item \textbf{Dual-functional radar-communication (DFRC) aided tracking (DFRC-T)} \cite{liu2020radar}: In \cite{liu2020radar}, an EKF-based method is proposed where the DFRC signals are used for probing the target to avoid dedicated pilots. 
%	However, both the radar sensing and communication functionalities are likely to experience performance degradation when being applied to dynamic scenarios. 
	The state variables used are different from those in EKF-T. The vehicle is also assumed to move in an ideal straight lane.

	\item \textbf{DL-based predictive beamforming (DL-PB)} \cite{mu2021integrated}: In \cite{mu2021integrated}, an FC network is designed to estimate the current angular parameter and calculate the beamforming angle at the next time slot based on the state evolution model established on the ideal straight lane. The echoes of DFRC signals are used as network input.
	
	\item \textbf{Vision-aided mmWave beam prediction (V-A-BP)} \cite{alrabeiah2020millimeter}\footnote{It is worth mentioning that V-A-BP and MMFF-Net are not intended to solve the same type of problem. Therefore, adjustments are made to the reproduction of the V-A-BP under our evaluation metrics in Section \ref{Performance}, which may incur some fairness issues.}: In \cite{alrabeiah2020millimeter}, the proactive beamforming task is degenerated to an image classification one and RGB images are utilized to predict the optimal beam from a pre-defined codebook. The images where the vehicle is at similar positions are classified into one category. Consequently, RSU uses the same beam to communicate with the vehicle at similar positions, which inevitably leads to the deviation of beam alignment.
	
	\item \textbf{Image-aided predictive beamforming (I-PB)}: A uni-modal data based scheme that adopts images to predict the vehicle position is designed. {\color{black}The I-PB scheme is designed to explore the role and effect of the 2D-visual feature for vehicle position prediction. Therefore, apart from the absence of data from the other two modalities and their respective NN modules, the image parameters, data processing methods, and network architecture of I-PB are identical to those of MMFF-Net.}
	
	\item \textbf{Depth map-aided predictive beamforming (D-PB)}: A uni-modal data based scheme that adopts depth maps to predict the vehicle position is designed.  {\color{black}The D-PB scheme is designed to explore the role and effect of the distance feature for vehicle position prediction. Similar with I-PB, the only difference between D-PB and MMFF-Net is the absence of CSI and images.}
	
	\item \textbf{CSI-aided predictive beamforming (C-PB)}: A uni-modal data based scheme that adopts CSI to predict the vehicle position is designed. Note that the mapping between CSI and the vehicle's absolute coordinates is difficult to fit through NN since CSI represents the compressed electromagnetic environment feature and the vehicle's position is just one of the contributing factors. Therefore, a temporal-difference prediction scheme is designed. That is, the NN predicts the displacement of vehicles at adjacent time slots through the potential angle variation feature in CSI sequences. {\color{black}The CSI data and its processing methods have not been changed.}

	\end{itemize}	
	
%	\vspace{-0.2cm}
	\section{Performance Evaluation}
	\label{Performance}
	In this section, extensive simulation results are presented to verify the effectiveness and superiority of the proposed scheme over the benchmarks. All the simulation results presented are on the average of $300$ independent realizations.
	\subsection{Angle Prediction}
	\label{angle perf}
\begin{figure*}[!t]
		\centering
		\includegraphics[width=1\textwidth]{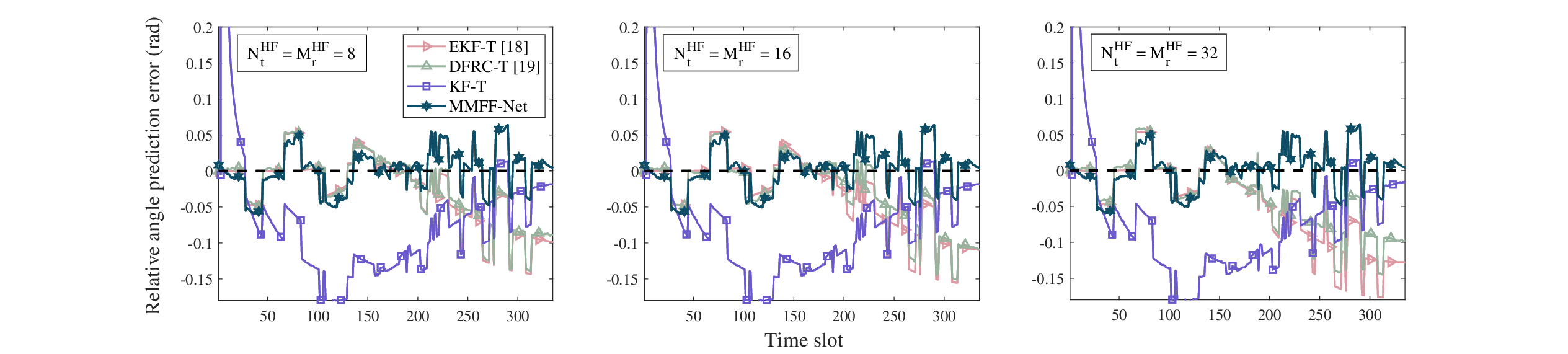}
		\caption{\color{black}Angle prediction performance comparisons among MMFF-Net, KF-T, EKF-T \cite{shaham2019fast}, and DFRC-T \cite{liu2020radar} schemes.}
		\label{relative_angle}
	\end{figure*}

	First, the angle prediction performances of the proposed MMFF-Net and benchmarks are given in Fig. \ref{relative_angle}. To present the results more clearly, the relative errors are presented instead of the absolute prediction results. The EKF-T and DFRC-T periodically use the received radar echoes at BS as the measurement signal vector, which causes excessive signaling overhead. However, the MMFF-Net does not consume extra communication overhead to detect vehicle's motion status periodically, thereby saving extensive spectrum resources.
%	The number of transmit and receive antennas affects the angle tracking accuracy. The angle prediction errors of EKF-T and HTP-PB get smaller as $N_{\rm{t}}^{\rm{HF}}$ and $M_{\rm{r}}^{\rm{HF}}$ increase since higher dimensional signal echoes provide more refined calibrations to the predicted values.
%	For fair comparisons, the variances of measurement noises are set to be consistent with that given in \cite{shaham2019fast,liu2020radar}.
	The angle tracking error of the KF-T is large due to its inapplicability in the non-linear motion model. The performance of DFRC-T is slightly better than that of EKF-T. The minimum angle prediction error of DFRC-T is $3.75 \times 10^{-2}$ when $N_{\rm{t}}^{\rm{HF}}=M_{\rm{r}}^{\rm{HF}} = 8$ but it is still $65.93\%$ higher than that of MMFF-Net. {\color{black}While the EKF-T and DFRC-T methods can initially achieve relatively accurate prediction results during the early stages of the tracking process, their errors gradually accumulate over time, causing the predicted angles to deviate from the actual angle, as depicted in Fig. \ref{relative_angle}. This deviation ultimately leads to the failure of beam tracking, which will be discussed in the next subsection. By contrast, the MMFF-Net  consistently  maintains stable and accurate prediction performance throughout the tracking process.}
	
%可以只放32的
	
\begin{figure}[!t]
		\centering
		\includegraphics[width=0.97\linewidth]{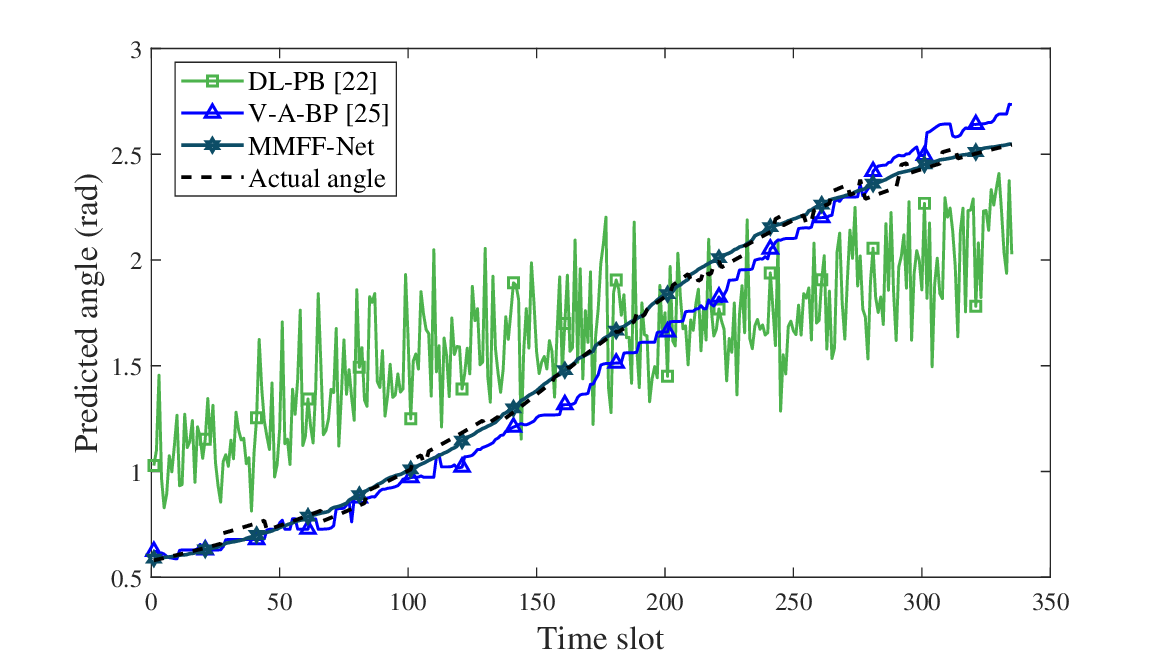}
		\caption{Angle prediction performance comparisons among MMFF-Net, DL-PB \cite{mu2021integrated}, and V-A-BP \cite{alrabeiah2020millimeter} schemes.}
		\label{v-d}
	\end{figure}

	Fig. \ref{v-d} shows the angle tracking performances of MMFF-Net, DL-PB \cite{mu2021integrated}, and V-A-BP schemes \cite{alrabeiah2020millimeter}. The hyper-parameters for the network design and fine-tuning given in \cite{mu2021integrated, alrabeiah2020millimeter} are adopted in our simulations. The equivalent angle prediction values corresponding to the beam prediction results of V-A-BP are obtained, which are finite discrete values due to the codebook with finite angular resolution. As can be seen in Fig.~\ref{v-d}, the V-A-BP can only complete rough angle tracking with relatively large errors, especially when the vehicle approaches the RSU. The DL-PB can hardly realize effective tracking of the vehicle since it merely relies on limited received signals and simple NN architecture.
%	  N_t^{\it{HF}}= M_r^{\it{HF}}= 8
	
	To further illustrate the necessity and superiority of multimodality fusion, the angle prediction performance of MMFF-Net and uni-modal schemes is presented in Table \ref{tab:inner comparison}. The average angle prediction errors of C-PB and I-PB are larger than DFRC-T and EKF-T, which indicates that merely relying on the angular feature and 2D-visual feature is not sufficient for accurate vehicle tracking. Among the uni-modal schemes, in terms of angle prediction performance, the D-PB scheme performs the closest to MMFF-Net but its average prediction error is still $46.02\%$ higher and the standard deviation of error is $45.70\%$ higher. Furthermore, when the vehicle approaches RSU, the angle prediction error of D-PB is $1.70 \times 10^{-2}$ and $13.33\%$ higher than that of MMFF-Net, demonstrating its weaker tracking ability for fast-varying angles. The performance gain brought by the proposed AWLN module can be reflected through the reduction in mean and standard deviation of angle prediction error, highlighting the significance of weight allocation among distinct multi-modal features.
	
	From the discussions above, it can be concluded that the superior performance of MMFF-Net is attributed to the efficient extraction of diverse behavioral features from multi-modal data that complement each other and reflect the vehicle's motion parameters more comprehensively and the thoughtful design of feature extraction and fusion mechanisms.  Despite the increased complexity, MMFF-Net is first trained offline and then deployed on high-performance RSU for online operation. Therefore, the increased complexity does not pose significant difficulties in MMFF-Net's practical application.

	\begin{table}[!htp]
		\setlength{\abovecaptionskip}{0.1cm} 
        \renewcommand\arraystretch{0.9} 
		\centering
		\caption{Angle prediction performances of MMFF-Net, MMFF-Net w/o AWLN module, and other uni-modal schemes.}
		\label{tab:inner comparison}
	
			\begin{tabular}{c|c|c}
				\toprule[0.35mm]
				\textbf{Schemes}	 &
				\makecell[c]{\textbf{Mean of angle}\\\textbf{prediction error (rad)}}	 & 	\makecell[c]{\textbf{Standard deviation of angle}\\ \textbf{prediction error (rad)}} \\
				\midrule[0.2mm]
				I-PB	& $4.24 \times 10^{-2}$ & $4.08 \times 10^{-2}$  \\ 
				\midrule[0.2mm]
				D-PB	& $3.30 \times 10^{-2}$ & $2.71 \times 10^{-2}$  \\ 
				\midrule[0.2mm]			
				C-PB	& $7.15 \times 10^{-2}$ & $5.70 \times 10^{-2}$  \\ 
%				\hline			
%			    EKF-T	& 0.0540(32) 0.0432(16) 0.0385(8) & 0.0495(32) 0.0419(16) 0.0376(8)  \\ 
%			    \hline			
%			    DFRC-T  & 0.0451(32) 0.0382(16) 0.0375(8) & 0.0382(32) 0.0386(16) 0.0348(8) \\ 
%				\hline			
%				HTP-PB	& 0.0225(32) 0.0223(16)  0.0248(8) & 0.0738(32) 0.0730(16) 0.0766(8) \\ 
%				\hline			
%				MMFF_w_noise	& 0.0254 & 0.0208 \\ 
			\midrule[0.2mm]		
			\makecell[c]{\textbf{MMFF-Net}\\  \textbf{w/o AWLN}}	& \textbf{$2.28 \times 10^{-2}$} & \textbf{$1.90 \times 10^{-2}$} \\ 
			\midrule[0.2mm]		
			\textbf{MMFF-Net}	& \textbf{$2.26 \times 10^{-2}$} & \textbf{$1.86 \times 10^{-2}$} \\ 
			    \bottomrule[0.35mm]
		\end{tabular}	
	\end{table}

	\subsection{Achievable Rate}
	\label{rate any}
\begin{figure*}[!t]
	\centering
	\includegraphics[width=1\linewidth]{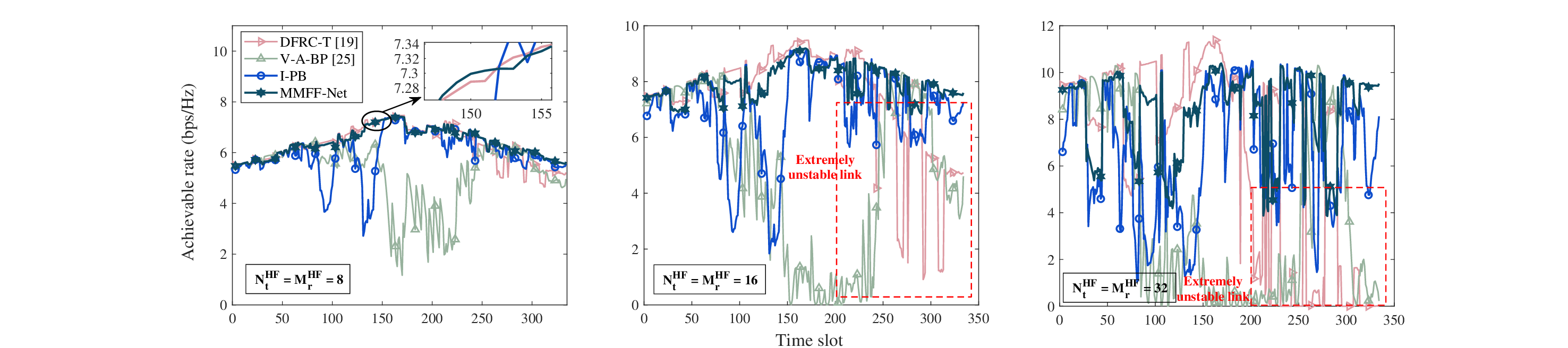}
	\caption{\color{black}Achievable rate performance comparisons among MMFF-Net, DFRC-T \cite{liu2020radar}, V-A-BP \cite{alrabeiah2020millimeter}, and I-PB schemes.}
	\label{rate}
\end{figure*}

	In Fig. \ref{rate}, the achievable rate comparisons between MMFF-Net and several benchmark schemes are shown. The transmit power $p_n$ is set to $15$dB. In the simulation scenario, the vehicle travels from one side of the RSU to the other, and the distance between them decreases first and then increases. Consequently, most of the achievable rates increase first, reach the maxima when the vehicle passes RSU, and then decrease, as can be observed in Fig. \ref{rate}. Furthermore, the average achievable rates increase as $N_{\rm{t}}^{\rm{HF}}$ and $M_{\rm{r}}^{\rm{HF}}$ increase thanks to the corresponding increased array gain. As shown in Fig. \ref{rate}, the MMFF-Net and DFRC-T schemes achieve similar performances in 8-antenna array case, both outperforming the I-PB. In 16-antenna and 32-antenna array cases, the advantage of MMFF-Net over I-PB and DFRC-T becomes notable especially in the latter half of the tracking process. The average achievable rates of MMFF-Net, I-PB, D-PB, and DFRC-T are $7.97$bps/Hz, $6.96$bps/Hz, $6.77$bps/Hz, and $7.48$bps/Hz in 16-antenna array case, and  $8.32$bps/Hz, $6.64$bps/Hz, $6.20$bps/Hz, and $6.30$bps/Hz in 32-antenna array case, respectively. It is worth noting that the V2I link that DFRC-T establishes is extremely unstable in the latter half of the tracking process since the linearization assumption and the approximation of the model in EKF algorithm causes error accumulation issue when dealing with highly nonlinear models, which severely limits its practical application effectiveness. However, MMFF-Net can effectively avoid this issue thanks to its powerful capability in handling nonlinear relationships and its characteristic of not requiring an explicit prior model. In the latter half of the tracking process, the average achievable rates of DFRC-T are $6.80$bps/Hz and $3.52$bps/Hz in 16 and 32-antenna array cases, while those of MMFF-Net are $8.05$bps/Hz and $8.52$bps/Hz.

	Fig.~\ref{rate} also presents the effect of vehicle drifting on achievable rate performance. The drifting behavior is realized by randomly generating a trajectory from discrete points on five parallel lanes, which is an abrupt event without any sign. Therefore, neither the uni-modal schemes nor the proposed scheme can predict such an event. All schemes experience a notable rate decrease due to misalignment in drifting, as observed in Fig.~\ref{rate}.
	
	It is noteworthy that adding antennas does not necessarily ensure an increasing achievable rate. Despite the high array gain, a larger array generates a narrower beam, which is likely to miss the target vehicle and thus leads to a larger beam misalignment probability. As shown in Fig. \ref{rate}, the achievable rates of the DFRC-T, I-PB, and MMFF-Net for the 16-antenna array case are higher than those of 8-antenna array case since their angle prediction accuracy can meet the alignment accuracy demand of 16-antenna array case. However, the angle prediction results of all schemes are not accurate enough to align the narrow beam generated by the 32-antenna array when the vehicle drifts, thereby causing the 32-antenna array case to perform worse than the 16-antenna array case in terms of achievable rates when the vehicle drifts.

	\subsection{Outage Probability}
	To evaluate the improvement of the proposed proactive beamforming scheme on the system's outage capacity, we analyze the outage probability for different proactive beamforming approaches. Let $P(R_{\rm{T}})$ be the outage probability when $R_{\rm{T}}$ is the minimum required achievable rate, $R_n$ be the achievable rate at the $n$-th time slot, $N(R_{\rm{T}})$ be the total number of time slots that meet the minimum achievable rate demand, and $N$ be the total number of time slots, then
	\begin{subequations}
		\label{outage}
		\begin{align}
			& P(R_{\rm{T}})  = 1 - \frac{N(R_{\rm{T}})}{N}  \label{o1}\\
			& N(R_{\rm{T}}) = \sum_{n=0}^{N}{\chi(R_n,R_{\rm{T}})}  \label{o2}\\
			& \chi(R_n,R_{\rm{T}})=\left\{
			\begin{array}{cl}
				1 &,~R_n  \geq R_{\rm{T}} \\
				0 &,~\mathrm{otherwise}
			\end{array} \right.
			\label{o3}
			
		\end{align}
	\end{subequations}

	\begin{figure*}[!t]
	\centering
	\includegraphics[width=1\textwidth]{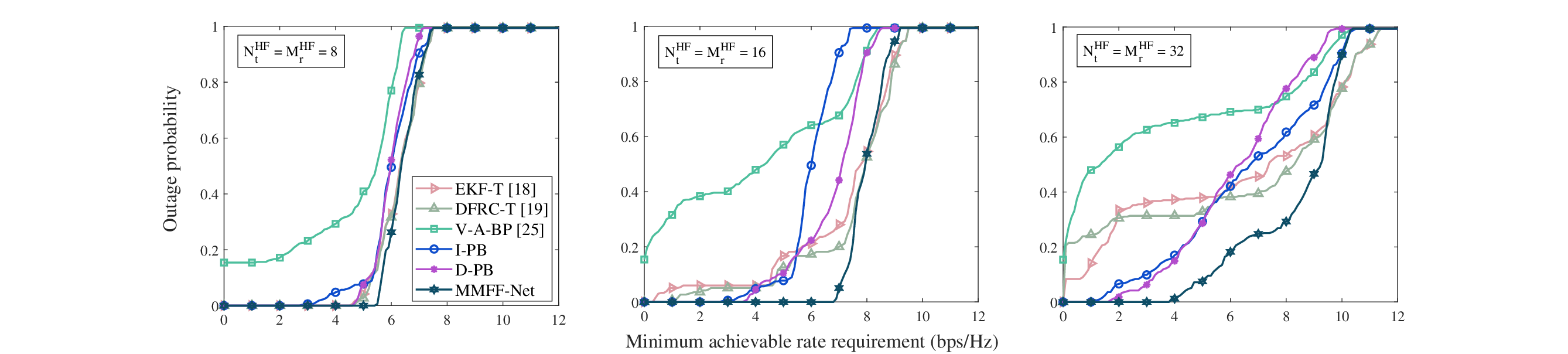}
	\caption{{\color{black}Comparisons of outage probability performance among MMFF-Net, EKF-T \cite{shaham2019fast}, DFRC-T \cite{liu2020radar}, V-A-BP \cite{alrabeiah2020millimeter}, D-PB, and I-PB schemes.}}
	\label{out}
    \end{figure*}

	Fig.~\ref{out} demonstrates the proposed scheme has a lower outage probability than other schemes, with the superiority becoming evident as $N_{\rm{t}}^{\rm{HF}}$ and $M_{\rm{r}}^{\rm{HF}}$ increase. As discussed in Section \ref{rate any}, a larger antenna does not necessarily lead to an increasing achievable rate since the narrower beam is more likely to miss the target vehicle. This conclusion can also be confirmed by the outage probability of the communication system. As shown in Fig. \ref{out}, the outage probability of the 32-antenna array case is larger than that of 16-antenna array case when using the MMFF-Net, under a minimum achievable rate requirement of $3.80$-$7.50$bps/Hz. DFRC-T achieves slightly higher achievable rates when the vehicle is very close to the RSU thanks to the abundant measurements with large-size antennas. This leads to a lower outage probability when the required minimum rate is higher than $8.00$bps/Hz and $9.50$bps/Hz in 16 and 32-antenna array cases.  However, for the overall tracking process, the outage probability is minimized when adopting MMFF-Net. MMFF-Net achieves a reduction of $1.14\%$, $5.14\%$, and $17.39\%$ in the outage probability compared to DFRC-T in 8, 16, 32-antenna array cases,  $5.46\%$, $20.70\%$, and $15.69\%$ compared to I-PB, and $5.48\%$, $12.55\%$, and $19.82\%$ compared to D-PB.
%	lower than that when DFRC-T for $23.68\%$, $70.53\%$, and $80.87\%$ of the time and in 8, 16, 32-antenna array case, respectively. and keep the V2I link as stable as possible when the vehicle drifts.
%	\vspace{-0.1cm}
	
	\subsection{Robustness Testing Against Adverse Environmental Conditions}
	\begin{table*}[!htp]
		\setlength{\abovecaptionskip}{0.1cm} 
		\renewcommand\arraystretch{0.7} 
		\centering
		\caption{Achievable rate performance of MMFF-Net confronted with adverse environmental conditions.}
		\label{tab:innernoise comparison}
		
		\begin{tabular}{>{\centering}m{4.9cm}|c|c|c}
			\toprule[0.35mm]
			\makecell[c]{\textbf{Schemes} \\ $(N_{\rm{t}}^{\rm{HF}},M_{\rm{r}}^{\rm{HF}})=(8,16,32)$} 	 &
			{\textbf{Mean of achievable rates}}	 & 	{\textbf{Maximum of achievable rates}} & {\textbf{Minimum of achievable rates}}\\	 
			\midrule[0.2mm]
			MMFF-Net 	& (6.39, 7.97, 8.32) & (7.41,  9.19,  10.39) & (5.48,  6.82,  3.86) \\
			
			\midrule[0.2mm]		
			\makecell[c] {MMFF-Net confronted with \\ adverse environment conditions}	& (6.11, 7.14, 6.92)  & (7.24, 8.51, 9.27) & (5.26, 5.35, 2.86) \\
			\bottomrule[0.35mm]
		\end{tabular}	
	\end{table*}
	The MMFF-Net system may encounter scenarios with varying degrees of errors in multi-modal data due to adverse environmental conditions. To assess the model's performance and robustness in such real-world scenarios, we manually introduce noise to the testing set simulating sensing device errors caused by environmental interference. Firstly, we reduce the RGB image brightness to simulate poor lighting conditions. Secondly, we introduce channel estimation errors to the CSI to simulate a non-ideal communication environment. The errors are measured using average normalized mean squared error (NMSE) and are set to $-15$dB. The NMSE is defined by ${\rm NMSE} = \frac{\Vert \hat{\mathbf{H}}_n - \mathbf{H}_n \Vert_{\rm{F}}^2}{\Vert \mathbf{H}_n \Vert_{\rm{F}}^2}$, with $\hat{\mathbf{H}}_n$ representing the CSI with errors. Thirdly, we increase the variance of the Gaussian noise added to the depth map to $1$ to simulate distance measurement errors due to weather factors. Then, the testing dataset is used to test the robustness of the MMFF-Net model pre-trained with the normal training set and the results are shown in Table \ref{tab:innernoise comparison}. The MMFF-Net inevitably experiences performance degradation and its average achievable rates are lower than those of MMFF-Net in normal environment condition but are $9.84\%$ and $4.16\%$ higher than those of DFRC-T and I-PB in 32-antenna array case. Therefore, the pre-trained MMFF-Net model exhibits robustness in maintaining a reliable V2I communication link even in the presence of environmental interference, demonstrating its reliable network architecture as well as strong feature extraction and fusion capabilities.

% max min average table}

%\begin{table*}[!htp]
%	\renewcommand\arraystretch{1.3} 
%	\centering
%	\caption{{\color{black}Angle tracking performances of MMFF-Net and other uni-modal schemes.}}
%	\label{tab:innernoise comparison}
%	
%	\begin{tabular}{>{\centering}m{3.8cm}|c|c|c|c|c|c|c|c|c}
%	\toprule[0.35mm]
%		\textbf{Schemes}	 &
%		\multicolumn{3}{>{\centering}c|}{\textbf{Mean of achievable rates}}	 & 	\multicolumn{3}{>{\centering}c|}{\textbf{Maximum of achievable rates}} & \multicolumn{3}{>{\centering}c}{\textbf{Minimum of achievable rates}}\\
%		\midrule[0.2mm]
%		 $(N_{\rm{t}}^{\rm{HF}},M_{\rm{r}}^{\rm{HF}})=(8,16,32)$ & \textbf{(8, 8)} & \textbf{(16, 16)} & \textbf{(32, 32)} & \textbf{(8, 8)} & \textbf{(16, 16)} & \textbf{(32, 32)} & \textbf{(8, 8)} & \textbf{(16, 16)} &  \textbf{(32, 32)}  \\
%%		 \midrule[0.35mm]
%%		 EKF-T \cite{shaham2019fast}	& 6.3453 & 7.8509 & 8.1663 & 7.4076 & 9.1071 & 10.2998 & 5.4631 & 6.4744 & 4.0695 \\
%		 
%		\midrule[0.35mm]
%		MMFF-Net	& 6.3453 & 7.8509 & 8.1663 & 7.4076 & 9.1071 & 10.2998 & 5.4631 & 6.4744 & 4.0695 \\
%		
%		\midrule[0.2mm]		
%		\makecell[c]{MMFF-Net Confronted with \\ Environmental Interference}	& 0.0230 & 0.0181 & 0  & 0.0236 & 0.0223 & 0 & 0  & 0.0236 & 0.0223 \\ 
%		\bottomrule[0.35mm]
%	\end{tabular}	
%\end{table*}

	\section{Conclusions}
	\label{VI}

	In this paper, we presented a novel proactive beamforming scheme for V2I links that leverages multi-modal sensing and communication integration. Our proposed scheme takes advantage of the complementary nature of multi-modal environment information and captures distinct features of the surrounding environment. To effectively extract and fuse the multi-modal features implied by the multi-modal data, we designed a novel MMFF-Net capable of pre-processing the multi-modal data with distinct data structures in the form of time series. To demonstrate the applicability of the proposed scheme to complex and dynamic scenarios, we constructed a dataset using the ViWi data-generation framework and enriched it by considering the vehicle's drifting behavior. We compared our proposed scheme with eight representative methods and revealed it can outperform all in terms of angle prediction accuracy, achievable rate, and outage performance. Moreover, our proposed scheme possesses the strongest robustness against vehicle drifting and environmental interference, demonstrating its practicality and effectiveness in complex dynamic scenarios. For future work, we aim to refine the CSI pre-processing at lower frequencies and extend the MMFF-Net to multi-vehicle scenarios.

\end{document}